\documentclass[a4paper,11pt]{article}
\pdfoutput=1 % if your are submitting a pdflatex (i.e. if you have images in pdf, png or jpg format)
\usepackage{jheppub} % for details on the use of the package, please see the JHEP-author-manual
\usepackage[T1]{fontenc} % if needed
\title{\boldmath Superpotentials, quantum parameter space and phase transitions
                         in $\mathcal{N}=1$ supersymmetric gauge theories}
\usepackage{graphicx}
\usepackage{amssymb}
\def\re{\mathop{\rm Re}\nolimits}
\def\im{\mathop{\rm Im}\nolimits}

\def\Log{\mathop{\rm Log}\nolimits}
\def\Tr{\mathop{\rm Tr}\nolimits}

\newcommand{\rmd}{\ensuremath{\mathrm{d}}}
\newcommand{\rmi}{\ensuremath{\mathrm{i}}}
\newcommand{\rme}{\ensuremath{\mathrm{e}}}
\author[a,1]{Gabriel \'Alvarez,\note{Corresponding author.}}
\author[a]{Luis Mart\'{\i}nez Alonso,}
\author[b]{and Elena Medina}

\affiliation[a]{Departamento de F\'{\i}sica Te\'orica II,
                               Facultad de Ciencias F\'{\i}sicas,
                               Universidad Complutense,
                               28040 Madrid, Spain}
\affiliation[b]{Departamento de Matem\'aticas,
                      Facultad de Ciencias,
                      Universidad de C\'adiz,
                      11510 Puerto Real, Spain}

\emailAdd{galvarez@fis.ucm.es}
\emailAdd{luism@fis.ucm.es}
\emailAdd{elena.medina@uca.es}
\abstract{We study the superpotentials, quantum parameter space and phase transitions
that arise in the study of large $N$ dualities between $\mathcal{N}=1$ SUSY $U(N)$ gauge
theories and string models on local Calabi-Yau manifolds. The main tool of our analysis is
a notion of spectral curve characterized  by a set of complex partial {}'t~Hooft parameters
and cuts given by projections on the spectral curve of minimal supersymmetric cycles of
the underlying Calabi-Yau manifold. We introduce a prepotential functional via a variational
problem which determines the complex density as an extremal constrained by the
period conditions. This prepotential is shown to satisfy the special geometry
relations of the spectral curve. We give a system of equations for the branch points
of the spectral curves and determine the appropriate branch cuts as Stokes lines of a suitable
set of polynomials. As an application, we use a combination of analytical and numerical methods
to study the cubic model, determine the analytic condition satisfied by critical one-cut spectral curves,
and characterize the transition curves between the one-cut and two-cut phases both
in the space of spectral curves and in the quantum parameter space.}
\begin{document}
\maketitle
\flushbottom
%%%%%%%%%%%%%%%%%%%%%%%%%%%%%%%%%%%%%%%%%%%%%%%%%%%%%%%%%%%%
%% INTRODUCTION %%%%%%%%%%%%%%%%%%%%%%%%%%%%%%%%%%%%%%%%%%%%%%%%
%%%%%%%%%%%%%%%%%%%%%%%%%%%%%%%%%%%%%%%%%%%%%%%%%%%%%%%%%%%%
\section{Introduction}
The aim of this paper is to apply the theory of  spectral curves to analyze the phase structure and the critical
processes arising in $\mathcal{N}=1$ SUSY $U(N)$ gauge theories with adjoint matter $\Phi$ obtained by
deforming the $\mathcal{N}=2$ theories by a tree-level superpotential $\Tr W(\Phi)$. Spectral curves with $s$ cuts
are associated to the classical vacua that break the gauge group $U(N)$ as a direct product of $s$ factors 
$U(N_1)\times \cdots \times U(N_s)$ where $N=N_1+\cdots+N_s$. These  spectral curves
arise as a consequence  of the large $N$ dualities between supersymmetric Yang-Mills theories
and string models on local Calabi-Yau manifolds $X$ of the form~\cite{CA01,DI02,DI022,HE07}
\begin{equation}
	\label{eq:cy}
 	W'(z)^2 + f(z)+u^2+v^2+w^2=0,
\end{equation}
where $W(z)$ and $f(z)$ are  polynomials 
\begin{equation}
	\label{pol}
	W(z) = \frac{z^{n+1}}{n+1} + t_n z^n + \cdots + t_1 z,
\end{equation}
\begin{equation}
	f(z) = b_{n-1} z^{n-1} + \cdots + b_0.
\end{equation}
 The corresponding
tree-level prepotential $\mathcal{F}$ can be characterized as a function of  the partial {}'t~Hooft
parameters $S_i$ by the special geometry relations
\begin{equation}
	\label{spe}
	S_i = \oint_{\mathbb{A}_i} \Omega,
	\quad
	\frac{\partial\mathcal{F}}{\partial S_i} = \oint_{\mathbb{B}_i}  \Omega,
\end{equation}
where $\Omega$ is the holomorphic $(3,0)$ form in $X$
and $\mathbb{A}_i$ and $\mathbb{B}_i$ form a symplectic basis of three-cycles in $X$.
In turn, these three-cycles can be understood as fibrations of two-spheres over paths
 in the Riemann surface defined by the spectral curve
\begin{equation}
	\label{eq:c1}
	y^2 = W'(z)^2 + f(z).
\end{equation}
Integration of $\Omega$ over the fibers reduces the integrals~(\ref{spe}) to integrals
of $y(z)\rmd z$ over the projections of the fibers onto the spectral curve.

In this paper we consider spectral curves $\Sigma$ determined from the following data:
\begin{enumerate}
	\item A set of $s$ pairs of branch points $a_i^{\pm}$ joined by $s$
	         finite disjoint cuts $\gamma_i$.
	\item A set of  $s$ nonzero complex numbers $S_i$
	         such that the branch of $y(z)$ with asymptotic behavior
		\begin{equation}
			\label{yin}
 			y(z) = W'(z) + \mathcal{O}(z^{-1}),\quad  z\rightarrow\infty,
		\end{equation}
	satisfies the period conditions
 	\begin{equation}
		\label{periods}
		\oint_{A_j}  y(z)  \rmd z = -4 \pi  \rmi  S_j,
	\end{equation}
	where $A_j$ is a counterclockwise contour encircling the cut $\gamma_j$.
\end{enumerate}
When we need to make explicit the fact that the spectral curves depend on both the cuts and  the partial {}'t~Hooft
parameters we denote these spectral curves by $\Sigma(\gamma,\mathbf{S})$, where
$\gamma = \gamma_1\cup\gamma_2\cup \ldots\cup \gamma_s$ and $\mathbf{S}=(S_1,\ldots,S_s)$. 

We emphasize that our analysis of spectral curves does not rely on random models
of matrices with eigenvalues constrained to lie on some path $\Gamma$ in the complex plane (holomorphic matrix models).
Our motivation is that a proper definition of these models  and their planar limit involves several deep subtleties~\cite{BI05}.  
For example, the saddle point solutions which provide the planar limit of the free energy exist only if  the path $\Gamma$ is such 
that the corresponding eigenvalue density is real and positive. This condition is trivially satisfied by hermitian models, 
where $\Gamma$  is the real line and the coefficients of $W(z)$ are real numbers,
but it represents  a quite non trivial requirement for general 
holomorphic matrix models. As a consequence,
the characterization of the cuts of the spectral curve  in terms of the support of the 
eigenvalue density of holomorphic matrix models  is a difficult problem. But the precise form of the cuts is obviously required  to  
analyze  critical processes such as cut splitting~\cite{HE07,FE03,CA03,HE08,MA10,AL10} and to study global features of the 
phase structure of the set of  spectral curves. 

In this paper we avoid any ambiguity by consistently using as cuts the projections onto
the spectral curve of minimal supersymmetric cycles
in the Calabi-Yau space $X$~\cite{BE95,KL96,SH99,GU00,GU00err}. These \emph{minimal cuts}
are characterized by the condition that the phase of $y(z) \rmd z$ is constant along each cut.
Minimal cuts are the natural generalization of the spectral cuts of the asymptotic eingenvalue density
in hermitian matrix models~\cite{BI05,FE04}.

One of the main reasons to introduce  matrix models in the study of gauge/string dualities is that the  planar limit of the
matrix model  free energy  provides the  tree-level prepotential  $\mathcal{F}$~\cite{DI02}.  However we will show 
that for any spectral curve the same expression of the prepotential
 \begin{equation}
	\label{pre00}
	\mathcal{F}
	=
	\int_{\gamma} W(z) \rho(z)  |\rmd z| 
	-
	\frac{1}{2} \int_{\gamma}  |\rmd z| \int_{\gamma}  |\rmd z'| \Log(z-z')^2 \rho(z) \rho(z')
\end{equation}
naturally appears  from a variational characterization of the complex density $\rho(z)$ defined by
\begin{equation}
	\label{mes1}
	\rho(z) | \rmd z| = y(z_+) \frac{ \rmd z}{2 \pi\rmi}=-y(z_-) \frac{ \rmd z}{2 \pi\rmi},
	\quad
	z\in\gamma,
\end{equation}
and constrained by the period conditions~(\ref{periods}), which in terms of $\rho(z)$ are
\begin{equation}
	\label{eq:den}
	\int_{\gamma_j}  \rho(z) | \rmd z| =  S_j.
\end{equation}
Incidentally, the subindices $z_+$ and $z_-$ in~(\ref{mes1})  refer to the one-sided limits of the
corresponding function $y(z)$ on $\gamma$, and in~(\ref{pre00}) the logarithm $\Log(z-z')^2 $
has to be understood as
\begin{equation}
	\Log(z-z')^2 = \log(z_+-z')+\log(z_--z'),\quad z,z'\in \gamma,
\end{equation}
for consistently chosen branches of $ \log(z-z')$.
The prepotential functional~(\ref{pre00}) is independent of the precise form of the cuts as long as
they remain in their respective homology classes in the complex plane with all the branch points deleted. 

The fundamental application of the prepotential $\mathcal{F}$  as a function of the coefficients $\mathbf{t}$
of $W(z)$ and the partial {}'t~Hooft parameters $\mathbf{S}$ is the determination of  the vacuum expectation
values (vevs)
\begin{equation}
	\label{vevs}
	u_k=\frac{1}{k}\langle \Tr \Phi^k \rangle,\quad \mathcal{S}_i=\langle S_i \rangle,\quad k=1,\ldots,n,\, \,i=1,\ldots,s,
\end{equation}
in the vacuum states $|\mathbf{k}\rangle$ labelled by $ \mathbf{k}=(k_1 k_2\cdots k_s),\,(k_i=1,\ldots,N_i)$,
which correspond to the  broken gauge  group $U(N_1)\times \cdots \times U(N_s)$. Thus~\cite{CA01,DI02}
if we introduce the  superpotential  $W_{\rm eff}({\bf t},{\bf S},\Lambda^2)$  by
\begin{equation}
	\label{supp}
	W_{\rm eff}=\sum_{i=1}^s N_i\,\frac{\partial \mathcal{F}}{\partial S_i}+ S\log \Lambda^{2 N},
\end{equation}
where 
\begin{equation}
	S = \sum_{i=1}^s S_i
\end{equation}
and $\Lambda$ is the nonperturbative scale in the $\mathcal{N}=2$ gauge theory, the vevs $\mathcal{S}_j$
are the solutions of the  field equations 
\begin{equation}
	\label{fie}
	\frac{\partial W_{\rm eff}}{\partial S_i}=0,\quad i=1,\ldots,s,
\end{equation}
and
\begin{equation}
	\label{vevs2}
	u_k = \frac{\partial W_{\rm low}}{\partial t_k}, \quad
	\sum_{i=1}^s \mathcal{S}_i=\frac{\partial W_{\rm low}}{\partial \log \Lambda^{2N}},
\end{equation}
where
\begin{equation}
	\label{low}
	W_{\rm low}({\bf t},\Lambda^2)
	=
	W_{\rm eff}({\bf t},\mathcal{S}_1({\bf t},\Lambda^2),\ldots,\mathcal{S}_s({\bf t},\Lambda^2),\Lambda^2),
\end{equation}
is the low-energy superpotential. The  logarithm in~(\ref{supp}) is only defined modulo $2\pi\rmi$  so that
$\mathcal{S}_i$ and  $W_{\rm low}$ are multivalued functions of $\Lambda^2$.  An important problem
is the characterization of the quantum parameter space  $\mathcal{M}_{\rm q}$~\cite{FE03b,FE03,FE03d}
on which  all these functions are single-valued.  For a fixed tree-level superpotential $W(z)$ of degree $n+1$
there is a decomposition of   the quantum parameter space 
\begin{equation}\label{dec}
	\mathcal{M}_{\rm q}=\mathcal{M}_{\rm q}^{(1)}\cup\cdots\mathcal{M}_{\rm q}^{(n)}
\end{equation}
into sectors $\mathcal{M}_{\rm q}^{(s)}$ corresponding to vacua with a fixed number $s$ of factors of the
broken gauge group.  Each point $( \mathcal{S}_1,\ldots,\mathcal{S}_s)\in \mathcal{M}_{\rm q}^{(s)}$
determines a class of $s$-cut spectral curves with  partial {}'t~Hooft parameters equal to $\mathcal{S}_i$.
The characterization of the subsets of spectral curves with minimal cuts in the sectors $\mathcal{M}_{\rm q}^{(s)}$
and their possible interpolations by smoothly varying the parameters of the theory is also an important issue.

In this work we  use a combination of analytic and numerical methods to study the phase structure
and the phase transitions in the space of  spectral curves with minimal cuts for a given polynomial $W(z)$.
These phases are labelled by the number $s$  of cuts  and are described by manifolds with points
parametrized by the partial {}'t~Hooft parameters. Moreover, using the correspondence between points of
the quantum parameter space  and spectral curves we translate our analysis of the phase structure
of spectral curves to the quantum parameter space of $\mathcal{N}=1$ SUSY $U(N)$ gauge theories.

The layout of this paper is as follows. In section~\ref{sec:sc} we explain our method
to characterize spectral curves and minimal cuts, briefly review some results concerning
the classical limit, and make precise the notion of critical spectral curves.
Section~\ref{sec:prep} is devoted to the study of the prepotential  associated
to a spectral curve via a variational characterization of the complex density;
then we consider the corresponding superpotential and the characterization of quantum vacua
in terms of solutions of the field equations. In section~\ref{sec:crit}  we study critical spectral
curves and, in particular, we analyze the phase transition corresponding to the splitting of one cut  in the
quantum parameter space. Sections~\ref{sec:onecc} and~\ref{sec:twocc} contain our numerical and
analytic study of the spectral curves and the quantum parameter space  for the cubic model.
Using our analytic condition (derived in section~\ref{sec:cc}) satisfied by critical spectral curves,
we characterize  the transition curves  between the one-cut and 
two-cut phases in both the space of spectral curves and the quantum parameter space.
The paper ends with a brief summary and we defer to two appendixes some technical proofs.
%%%%%%%%%%%%%%%%%%%%%%%%%%%%%%%%%%%%%%%%%%%%%%%%%%%%%%%%%%%%
\section{Spectral curves and minimal cuts \label{sec:sc}}
%%%%%%%%%%%%%%%%%%%%%%%%%%%%%%%%%%%%%%%%%%%%%%%%%%%%%%%%%%%%
In this section we discuss the characterization of spectral curves (\ref{eq:c1}) with minimal cuts.
We denote by $a_1,\ldots,a_n$ the critical points of $W(z)$ 
\begin{equation}
	W'(z) = (z-a_1)\cdots(z-a_n).
\end{equation}
We will assume that the roots of $y^2(z)$ are either simple or double, and denote the double roots
by $\alpha_1,\ldots,\alpha_r$. Hence the function $y(z)$ for an $s$-cut spectral curve can be written as
\begin{equation}
	\label{y1}
	y(z) = h(z) w(z),
\end{equation}
where
\begin{equation}
	h(z) = \prod_{l=1}^{r} (z-\alpha_l),
	\quad
	w(z) = \sqrt{\prod_{m=1}^{s} (z-a_m^{-})(z-a_m^{+})},
\end{equation}
$r+s=n$, and where the branch of $w(z)$ is fixed by
\begin{equation}
	w(z)\sim z^{s},\quad  z\rightarrow\infty.
\end{equation}
But according to~(\ref{yin}) the factor $h(z)$ in~(\ref{y1}) is given by
\begin{equation}
	\label{ache}
	h(z) = \left( \frac{W'(z)}{w(z)} \right)_{\oplus},
\end{equation}
where $\oplus$ stands for the sum of the nonnegative powers of the corresponding Laurent series at infinity.
Therefore the function $y(z)$ is completely determined by its branch points, the simple roots
$a_1^{\pm},\ldots,a_s^{\pm}$. We will often use the variables
\begin{equation}\label{bd}
\beta_i=\frac{a_i^{+}+a_i^{-}}{2},\quad \delta_i=\frac{a_i^{+}-a_i^{-}}{2}.
\end{equation}
%%%%%%%%%%%%%%%%%%%%%%%%%%%%%%%%%%%%%%%%%%%%%%%%%%%%%%%%%%%%
\subsection{Determination of  spectral curves with minimal cuts\label{sec:dec}}
%%%%%%%%%%%%%%%%%%%%%%%%%%%%%%%%%%%%%%%%%%%%%%%%%%%%%%%%%%%%
To determine the endpoints $a_1^{\pm},\ldots,a_s^{\pm}$ of an $s$-cut spectral curve we
substitute~(\ref{y1}) into the left-hand side of~(\ref{eq:c1}) and identify the coefficients of 
$1,z,\cdots, z^{n+s-1}$ in both members. The remaining coefficients do not give independent
relations because
\begin{eqnarray}
	\label{sh1}
	\nonumber y(z)^2-W'(z)^2
 	& = & \Big(\frac{W'(z)}{w(z)}\Big)_{\ominus}
	         \Big[\Big(\frac{W'(z)}{w(z)}\Big)_{\ominus} w(z)^2 - 2 W'(z) w(z)\Big]\\
        & = & \mathcal{O}(z^{n+s-1}),\quad z\rightarrow \infty.
\end{eqnarray}
Thus we find $n+s$ equations which, however, involve the coefficients $b_0,\ldots,b_{n-1}$ of the
polynomial $f$. The $s$ additional independent relations follow from imposing
the period conditions~(\ref{periods}).

Incidentally, it will be useful later to note that as a consequence of~(\ref{yin}) and~(\ref{periods}),
the coefficient $b_{n-1}$ of $f(z)$ is given by
\begin{equation}
	b_{n-1} = -4\,S.
\end{equation}
Note also that implicit in the calculation of the endpoints for $s>1$  is the choice of the $s$ cuts
$\gamma_1,\ldots,\gamma_s$ connecting the unknown pairs of endpoints $a_1^{\pm},\ldots,a_s^{\pm}$.

In general this method leads to several families of solutions for the set of branch points $a_j^{\pm}$
as functions of the {}'t~Hooft parameters $\mathbf{S}$ which, in turn, determine several families
of spectral curves $\Sigma(\gamma,\mathbf{S})$. Moreover, the values of the branch points
depend only on the homology classes of the cuts
$\gamma_j$ in $\mathbb{C}\setminus\{a_1^{\pm},\ldots,a_s^{\pm}\}$.

\begin{figure}
  \begin{center}
  \includegraphics[width=8cm]{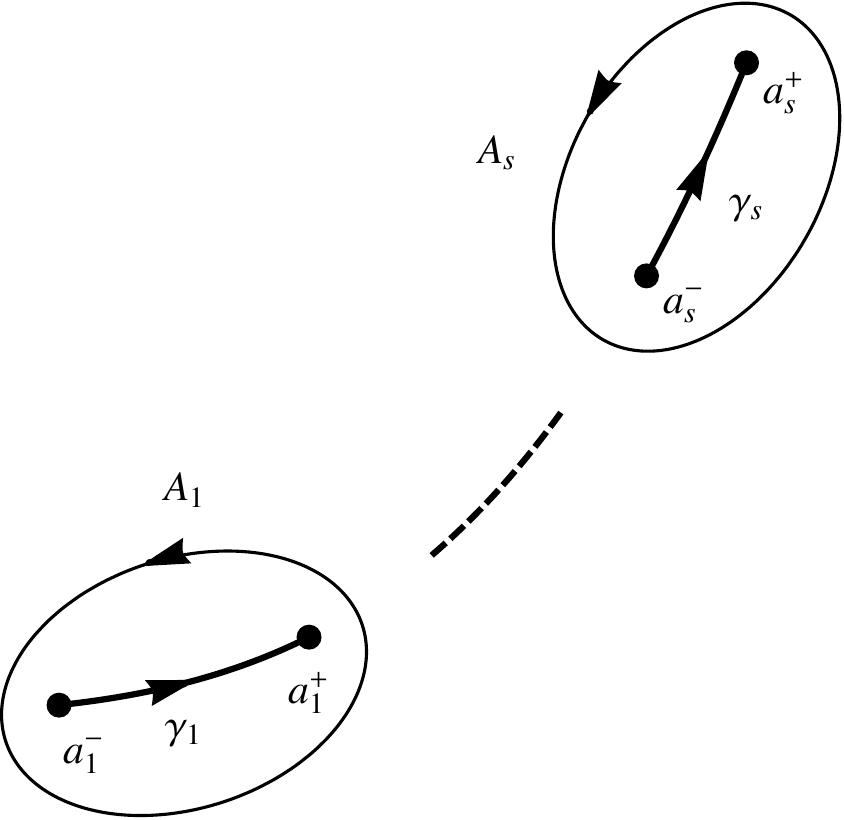}
  \end{center}
  \caption{Stokes lines $\gamma_i$ and contours $A_i$.\label{fig:ai}}
\end{figure}

The next step after determining the branch points is to characterize the set of values of $\mathbf{S}$
for which there exist minimal cuts $\gamma_j$ along which the phases of $y(z) \rmd z$
are constant (cf.~figure~\ref{fig:ai}), i.e., each cut $\gamma_j$ must be a Stokes line determined by
\begin{equation}
	\label{eq:st}
	\re G_j (z) = 0,\quad z\in \gamma_j,
\end{equation}
where
 \begin{equation}
	\label{mesn}
	 G_j(z)= \rme^{-\rmi\arg S_j} \int_{a_j^{-}}^z y(z'_+)  \rmd  z'.
\end{equation}
Note that although the integration path in~(\ref{mesn}) is left undefined,  due to the period
conditions~(\ref{periods}) the real parts of the functions $G_j(z)$ are single-valued.

Unfortunately, except for certain symmetric cases only a few general facts can be stated about
Stokes complexes~\cite{SI95}: since the endpoints $a_j^{\pm}$ are simple zeros of 
$y^2(z)$, three Stokes lines stem from each branch point forming equal angles $2\pi/3$,
and each of these Stokes lines does not make loops and ends either
at a different zero of $y(z)$ or at infinity. We anticipate that the study of critical configurations
requires the calculation of the complete Stokes graph, which comprises
both the Stokes lines~(\ref{eq:st}) stemming from the simple roots $a_j^{\pm}$
and the four Stokes lines stemming from each double root $\alpha_l$ (if they exist).
The calculation of a Stokes graph in a generic case has to rely on numerical methods.
%%%%%%%%%%%%%%%%%%%%%%%%%%%%%%%%%%%%%%%%%%%%%%%%%%%%%%%%%%
%%%%%%%%%%%%%%%%%%%%%%%%%%%%%%%%%%%%%%%%%%%%%%%%%%%%%%%%%%
\subsection{The Gaussian model}
%%%%%%%%%%%%%%%%%%%%%%%%%%%%%%%%%%%%%%%%%%%%%%%%%%%%%%%%%%%%%
The simplest example of spectral curves with minimal cuts is provided by the Gaussian model
\begin{equation}
	\label{gau}
	W(z) = \frac{z^2}{2},
\end{equation}
for which only one-cut spectral curves may arise.  Substituting $y^2=(z-a^-)(z-a^+)$
and $f(z)=-4S$ into~(\ref{eq:c1}) we get
\begin{equation}
	a^+ = -a^- =2\sqrt{S}.
\end{equation}
Then, for $z$ on the straight line segment $\gamma$ with endpoints $a^{\pm}$
\begin{equation}
	y(z_+) = \rmi \,|z^2-4 S|^{1/2} \rme^{\rmi\arg S/2}
\end{equation}
and therefore $\gamma$ is clearly a minimal cut.
%%%%%%%%%%%%%%%%%%%%%%%%%%%%%%%%%%%%%%%%%%%%%%%%%%%%%%%%%%%%%
\subsection{The classical limit\label{sec:sccl}}
%%%%%%%%%%%%%%%%%%%%%%%%%%%%%%%%%%%%%%%%%%%%%%%%%%%%%%%%%%%%%
A $s$-cut family $\Sigma(\gamma,\mathbf{S})$ of spectral curves is said to admit a  classical limit if
as $\mathbf{S}\rightarrow\mathbf{0}$  the $s$ cuts shrink to $s$ non-degenerate critical points of $W(z)$
\begin{equation}
	\label{limb}
	a_j^{\pm} \sim a_j ,\quad  (j=1,\ldots,s),
\end{equation}
while the double roots of $y^2(z)$ tend to the remaining critical points of $W(z)$
\begin{equation}
	\label{limbm}
	\alpha_l\sim a_{s+l},\quad  (l=1,\ldots,r),
\end{equation}
so that the family of  spectral curves degenerates into $y^2\sim W'(z)^2$. In this case
$|a_j^{+}-a_j^{-}|\ll|a_j-a_k|$ for all $k\neq j$ and then~(\ref{limb})--(\ref{limbm}) imply that for $z\in \gamma_j$
\begin{eqnarray}\label{34}
	y^2(z) & = & \prod_{l=1}^{r} (z-\alpha_l)^2
	                    \prod_{i=1}^{s} (z-a_i^{-})(z-a_i^{+})\nonumber\\
	           & \sim & W''(a_j)^2(z-a_j^{+})(z-a_j^{-}).
\end{eqnarray}
Then
\begin{equation}
	\frac{1}{2\pi\rmi}\,\int_{a_j^{-}}^{a_j^{+}} y(z_+) \rmd z
	 \sim
	\frac{W''(a_j)}{2\pi}
	\int_{a_j^{-}}^{a_j^{+}} \sqrt{(a_j^{+}-z)(z-a_j^{-})}\,\rmd z 
	= W''(a_j)\frac{(a_j^{+}-a_j^{-})^2}{16}.
\end{equation}
Hence from the period relations we get
\begin{equation}
	\label{bddd}
 	\frac{(a_j^{+}-a_j^{-})^2}{16}\sim \frac{S_j}{W''(a_j)},\quad  (j=1,\ldots,s).
\end{equation}
This means that the solutions of the cut endpoints equations in the classical limit is
\begin{equation}
	\label{epc}
	\beta_j\sim a_j,\quad \delta_j^2\sim \frac{4\,S_j}{W''(a_j)},\quad  (j=1,\ldots,s).
\end{equation}

Moreover, by adapting an argument used by Bilal and Metzger~\cite{BI05} in the context of
holomorphic matrix models, we can prove that in the classical limit these spectral curves
have minimal cuts which, to first order in $\mathbf{S}$, are the straight line segments with endpoints $a_j^{\pm}$.
According to~(\ref{34})
\begin{equation}
	\rho(z) = \frac{y(z_+)}{2\pi\rmi}
	               \frac{{\rm d}z}{|{\rm d}z|}
	           \sim \frac{|W''(a_j)|}{2\pi} \sqrt{|z-a_j^{+}||z-a_j^{-}|} \rme^{\rmi(\varphi_j+2\psi_j)}
\end{equation}	
where
\begin{equation}
	\rme^{\rmi\varphi_j} = \frac{W''(a_j)}{|W''(a_j)|},
	\quad
	\rme^{\rmi\psi_j} = \frac{a_j^{+}- a_j^{-}}{|a_j^{+}- a_j^{-}|}.
\end{equation}
Then setting $2\psi_j=\arg S_j-\varphi_j$,  the period conditions~(\ref{periods}) imply that the
segments $[a_j^{-}, a_j^{+}]$ are minimal cuts.
An alternative argument to see this property follows from an observation of Felder~\cite{FE04}:
as $\mathbf{S}\to\mathbf{0}$, the minimal cuts $\gamma_j$ are the level lines
\begin{equation}
	\re(\rme^{-\rmi\arg S_j} W(z) - \rme^{-\rmi\arg S_j} W(a_j)) =0.
\end{equation}
Hence, if $a_j$ is a non degenerate critical point of $W(z)$, any sufficiently small circle $C$ around $a_j$
intersects the level lines at four points. For small $\mathbf{S}$ the point $a_j$ splits into the two branch
points $a_j^{\pm}$ and, by continuity, they must be connected by a Stokes line~(\ref{eq:st})
(i.e., a minimal cut) inside the circle $C$ so that there are four Stokes lines leaving $C$.
%%%%%%%%%%%%%%%%%%%%%%%%%%%%%%%%%%%%%%%%%%%%%%%%%%%%%%%%%%%%
\section{Prepotentials, superpotentials and vacua \label{sec:prep}}
%%%%%%%%%%%%%%%%%%%%%%%%%%%%%%%%%%%%%%%%%%%%%%%%%%%%%%%%%%%%%
\subsection{Prepotentials associated to spectral curves}
\begin{figure}
  \begin{center}
  \includegraphics[width=8cm]{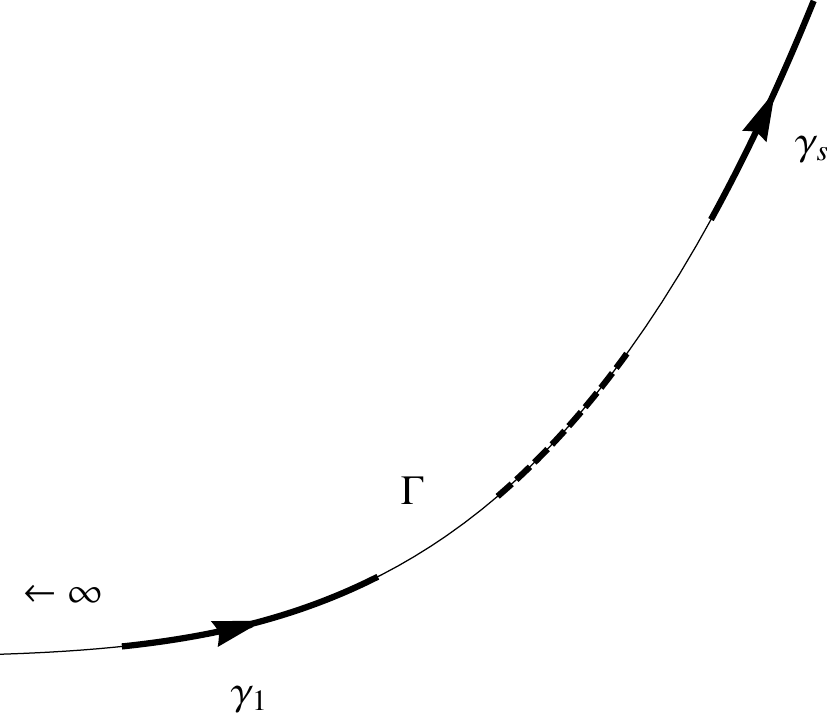}
  \end{center}
  \caption{Example of a semi-infinite path $\Gamma$ that contains the cuts~$\gamma_j$ and allows the
                construction of consistent determinations of the logarithms in the prepotential.\label{fig:pathgamma}}
\end{figure}

%%%%%%%%%%%%%%%%%%%%%%%%%%%%%%%%%%%%%%%%%%%%%%%%%%%%%%%%%%%%%
We first introduce a semi-infinite oriented path $\Gamma$  containing the cuts as shown in
figure~\ref{fig:pathgamma}, and such that for  each $z'$ in $\Gamma$ there exists an analytic branch of
$\log(z-z')$ as a function of $z$ in $\mathbb{C}$ minus the semi-infinite arc $\Gamma_{z'}$ of $\Gamma$
ending at $z'$, that verifies
\begin{equation}
    \label{loga}
      \log(z_+-z') + \log(z_--z') = \log(z'_+-z)+\log(z'_--z), \quad \mbox{for all $z\neq z'$ in $\Gamma$}.
\end{equation}
%%%%%%%%%%%%%%%%%%%%%%%%%%%%%%%%%%%%%%%%%%%%%%%%%%%%%%%%%%%
%%%%%%%%%%%%%%%%%%%%%%%%%%%%%%%%%%%%%%%%%%%%%%%%%%%%%%%%%%%
The property~(\ref{loga}) is essential for the consistency  of the following definition of  $\log(z-z')^2$
\begin{equation}
    \label{logq}
    \log(z-z')^2 = \log(z_+-z')+\log(z_--z'),\quad z,z'\in \Gamma,
\end{equation}
which is assumed in the expression of the prepotential~$\mathcal{F}$. 
It is easy to prove that for arcs $\Gamma$ with parameterization $z=z(\tau)$ such that at least one of the
functions $x(\tau)$ and $y(\tau)$ is strictly monotone,  the property~(\ref{loga}) is satisfied by the logarithmic
branches defined by
\begin{equation}
    \label{logbb}
    \log(z-z') = \log |z-z'| + \int_{\Gamma_{z,z'}} \frac{\rmd u}{u-z'},
    \quad z'\in \Gamma,
    \quad z\in \mathbb{C}\setminus \Gamma_{z'},
\end{equation}
where $\Gamma_{z,z'}$ is any path in $\mathbb{C}\setminus \Gamma_{z'}$ connecting $z'+|z-z'|$
to $z$. For example if $\Gamma$ is a real interval of the form $(-\infty, x_0]$  then~(\ref{logbb})
determines the principal branch of  $\log(z-z')$ and~(\ref{logq}) gives $\log(z-z')^2=2\log|z-z'|$.

Using hereafter these logarithmic branches, we consider the function
\begin{equation}
	\label{ge}
	g(z) = \int_{\gamma}\log(z-z') \rho(z') |\rmd z'|
\end{equation}
where $\rho(z)$ is the complex   density (\ref{mes1}). Note that both $y(z)-W'(z)$ and $-2g'(z)$
are analytic in $\mathbb{C}\setminus\gamma$,  vanish as $z\to\infty$ and, according to~(\ref{mes1}),
have the same jump on $\gamma$. Therefore
\begin{equation}
	\label{y}
	y(z) = W'(z)-2 g'(z),
\end{equation}
and using
\begin{equation}
	\label{rh}
	y(z_+) +y(z_-)=0,\quad z\in \gamma,
\end{equation}
we get
\begin{equation}
	\label{ds}
 	W'(z) - (g'(z_+)+g'(z_-)) = 0,\quad z\in\gamma.
 \end{equation}
Equation~(\ref{ds}) means that $W(z)-(g(z_+)+g(z_-))$ is constant on each connected piece
of $\gamma$ or, equivalently, that there are (not necessarily equal) complex numbers $L_i$ such that
\begin{equation}
	\label{s1}
	 W(z)-\left(g(z_+)+g(z_-)\right)=L_i,\quad z\in \gamma_i.
\end{equation}

A straightforward calculation using~(\ref{loga}) and~(\ref{logq}) shows that~(\ref{s1}) can be written as
the variational equation
\begin{equation}
	\label{var}
	\frac{\delta}{\delta\rho}
	\left[\mathcal{F}+\sum_{i=1}^s\,L_i \left(S_i\int_{\gamma_i}\rmd q(z)\right)\right] = 0,
\end{equation}
where
\begin{equation}
	\label{pre}
	\mathcal{F}
	=
	\int_{\gamma} W(z)\,  \rmd q(z)
	-
	\frac{1}{2} \int_{\gamma}  \rmd q(z)\int_{\gamma}  \rmd q(z') \,\Log(z-z')^2,
\end{equation}
is the prepotential functional and where
\begin{equation}
	\rmd q(z) = \rho(z) |\rmd z|.
\end{equation}
Thus, the variational equation~(\ref{var}) characterizes $\rho(z)$ as a (in general, local)
extremal density for $\mathcal{F}$ constrained by (\ref{eq:den}). It also follows that (cf.~Appendix~A)
\begin{equation}
	\label{eq:dfds}
	\frac{\partial\mathcal{F}}{\partial S_i} = L_i,
\end{equation}
so that we can express the superpotential in the form
\begin{equation}
	\label{supp2}
	W_{\rm eff}=\sum_{i=1}^s N_i\,L_i+ S\log \Lambda^{2 N}.
\end{equation}
Note that if we write (\ref{pre}) as
\begin{equation}
	\mathcal{F}
	=
	\int_{\gamma} \left( W(z)-\frac{1}{2}(g(z_+)+g(z_-))\right)\rmd q(z)
\end{equation}
and use again~(\ref{s1}), we obtain the usual alternative expression for the prepotential
\begin{equation}
	\label{pre1}
	\mathcal{F}
	=
	\frac{1}{2}\,\int_{\gamma}  W(z) \rmd q(z)  + \frac{1}{2} \sum_{i=1}^s S_i L_i.
\end{equation}
%%%%%%%%%%%%%%%%%%%%%%%%%%%%%%%%%%%%%%%%%%%%%%%%%%%%%%%%%%%%%
\subsection{The Gaussian model}
%%%%%%%%%%%%%%%%%%%%%%%%%%%%%%%%%%%%%%%%%%%%%%%%%%%%%%%%%%%%%
As an illustrative example we consider again the Gaussian model~(\ref{gau}) with the minimal cut  $\gamma$
given by the segment  $[-2\sqrt{S},2 \sqrt{S}]$. Let us take the path $\Gamma$ as the semi-infinite straight line
containing $\gamma$ and ending at $2\sqrt{S}$, and define $\log(z-z')$ for $z$ not in $\Gamma_{z'}$
according to~(\ref{logbb}). Then we have
\begin{equation}
 	\log(z_+-z') + \log(z_--z')
	=
	2 \log |z-z'| + \rmi \arg S,\quad z,z'\in\gamma.
\end{equation}
Hence if we parameterize $\gamma$ by $z(t) = 2 t |S|^{1/2} \rme^{\rmi\arg S/2}$ with $-1\leq t\leq 1$ we get
\begin{eqnarray}
	\mathcal{F} & = &\frac{4|S|^2 \rme^{\rmi 2 \arg S}}{\pi}
	                            \int_{-1}^{1} t^2\sqrt{1-t^2} \rmd t  - \rmi \frac{\arg S}{2} S^2
	\nonumber\\
                           &     & {}-\frac{4|S|^2 \rme^{\rmi 2 \arg S}}{\pi^2}
                                      \left[\int_{-1}^{1} \int_{-1}^{1} \sqrt{1-t^2}\sqrt{1-(t')^2} 
                                             \log |t-t'|\rmd t \rmd t' \right.
        \nonumber\\
                           &    & \qquad\qquad\qquad\qquad\left.{}+ \frac{\pi^2}{4} \log(2|S|^{1/2})\right]
	\nonumber \\
	\label{pr1}
	& = & \left(\frac{3}{4}-\frac{1}{2}\log S\right) S^2,
\end{eqnarray}
where $\log S=\log |S|+\rmi\arg S$.

Hence the superpotential is
\begin{equation}
	W_{{\rm eff}} = N S (1-\log S) + S \log \Lambda^{2N}.
\end{equation}
The field equations reduce to
\begin{equation}
	\log\Big(\frac{\Lambda^{2N}}{S^N}\Big) = 0,
\end{equation} 
and we get the $N$ vacua $|k \rangle$ of the $\mathcal{N}=1$ SUSY $U(N)$ gauge theory which
are characterized by the vevs
\begin{equation}
	\mathcal{S}^{(k)} = \zeta_k \Lambda^2,
\end{equation}
and the low-energy superpotentials
\begin{equation}
W_{{\rm low}}^{(k)}(\Lambda^2)= N \zeta_k \Lambda^2,
\end{equation}
where
\begin{equation}
	\zeta_k = \rme^{2\pi\rmi k/N},\quad k=1,\ldots,N.
\end{equation}
Thus the quantum parameter space $\mathcal{M}_{\rm q}$  is made of $N$ copies of the complex
$\Lambda^2$-plane connected at the point $\Lambda^2=0$.
%%%%%%%%%%%%%%%%%%%%%%%%%%%%%%%%%%%%%%%%%%%%%%%%%%%%%%%%%%%%%
\subsection{Vacua in the classical limit}
%%%%%%%%%%%%%%%%%%%%%%%%%%%%%%%%%%%%%%%%%%%%%%%%%%%%%%%%%%%%%
The leading approximation of $\mathcal{S}_j$ in the classical limit can be determined~\cite{CA01}.
In fact, it follows from the the asymptotic formula~(\ref{elesa}) of Appendix~A that
\begin{equation}
	L_j \sim W(a_j)-\sum_{k\neq j}S_k \log\Delta_{jk}^2+S_j \Big(1+\log\Big(\frac{W''(a_j)}{S_j}\Big)\Big),
\end{equation}
where $\Delta_{jk}\equiv a_j-a_k$. The corresponding classical limit for the superpotential is
\begin{equation}
	\label{supc}
	W_{{\rm eff}} \sim \sum_{j=1}^sN_j\,W(a_j)
	+
	\sum_{j=1}^s S_j
	\Big[\log\Big(\frac{W''(a_j)^{N_j}\prod_{i\neq j} \Delta_{ij}^{-2 N_i}\Lambda^{2 N}}{S_j^{N_j}}\Big)+N_j \Big],
\end{equation}
and the field equations in this approximation read
\begin{equation}
	\label{fiec}
	\log\Big(\frac{W''(a_j)^{N_j}\,\prod_{i\neq j} \Delta_{ij}^{-2 N_i}\,\Lambda^{2 N}}{S_j^{N_j}}\Big)=0,
\end{equation}
which give rise to $N_1\times N_2\times \cdots \times N_s$ vacua $|\mathbf{k}\rangle=|k_1 k_2\cdots k_s\rangle$
with approximate associated vevs~\cite{CA01}
\begin{equation}
	\label{vevesg}
	\mathcal{S}_j^{({\bf k})} \sim \rme^{2\pi\rmi k_j/N_j}
	W''(a_j)\,\Lambda^2\prod_{k\neq j}\Big(\frac{\Lambda}{\Delta_{jk}}\Big)^{2N_k/N_j},\quad k_j=1,\ldots,N_j.
\end{equation}
%%%%%%%%%%%%%%%%%%%%%%%%%%%%%%%%%%%%%%%%%%%%%%%%%%%%%%%%%%%%%
\subsection{Field equations in terms of  Abelian differentials \label{sec:abe}}
%%%%%%%%%%%%%%%%%%%%%%%%%%%%%%%%%%%%%%%%%%%%%%%%%%%%%%%%%%%%
The characterization of the derivatives of the prepotential with respect to
the partial {}'t~Hooft parameters is essential to determine the prepotential from~(\ref{eq:dfds}) and~(\ref{pre1}),
as well as to study phase transitions of spectral curves. In Appendix~A we prove that
\begin{equation}
	\label{prep}
	\frac{\partial \mathcal{F}}{\partial S_i}
	=
	L_i
	=
	W(z_i)-\int_{\gamma}  \Log(z-z_i)^2 \rmd q(z),\quad z_i\in\gamma_i,
\end{equation}
\begin{equation}
	\label{iij0}
	\frac{\partial^2 \mathcal{F}}{\partial S_i\partial S_j}
	=
	-\int_{\gamma}  \Log(z-z_i)^2 \frac{\partial  \rmd q(z)}{\partial S_j},\quad z_i\in\gamma_i.
\end{equation}
In this section we show that the second-order derivatives~(\ref{iij0}) can be expressed
in terms of Abelian differentials of the two-sheeted Riemann surface determined by~(\ref{eq:c1}). 
%%%%%%%%%%%%%%%%%%%%%%%%%%%%%%%%%%%%%%%%%%%%%%%%%%%%%%%%%%%
\begin{figure}
  \begin{center}
  \includegraphics[width=8cm]{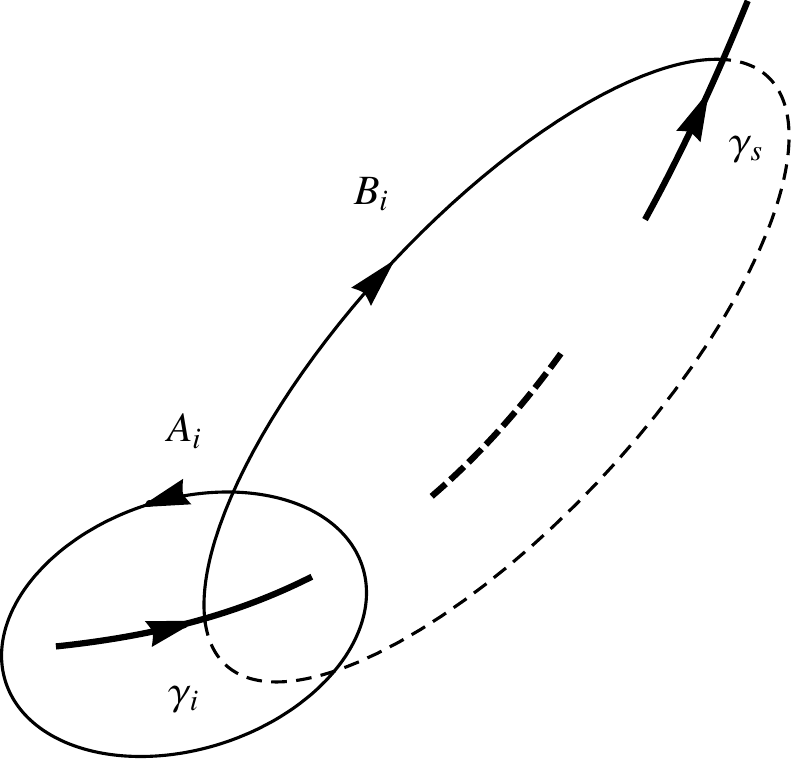}
  \end{center}
  \caption{Homology basis in the Riemann surface.\label{fig:basis}}
\end{figure}
%%%%%%%%%%%%%%%%%%%%%%%%%%%%%%%%%%%%%%%%%%%%%%%%%%%%%%%%%%%

We take the homology basis of cycles $\{A_1,\ldots,A_{s-1},B_1,\ldots,B_{s-1}\}$ as shown in figure~\ref{fig:basis},
and introduce the meromorphic differential  $\mathrm{y}(z)\rmd z$, where $\mathrm{y}(z)$
is the extension of the function (\ref{yin}) to the two sheets of the Riemann surface by means of the two branches
$y_1(z)=-y_2(z)=y(z)$, so that $y_2(z_-)=-y_2(z_+)=y_1(z_+)=-y_1(z_-)$.
Moreover, using~(\ref{y}) and~(\ref{s1}) we find that the $B$-periods are
\begin{equation}
	\label{perb}
	\oint_{B_j}\mathrm{y}(z)\rmd z = 2 (L_s-L_j),\quad j=1,\ldots,s-1.
\end{equation}

Let us  denote by $\rmd\phi_1,\ldots,\rmd\phi_{s-1}$ the canonical basis of normalized holomorphic differentials
\begin{equation}
	\label{nord}
	\oint_{A_i} \rmd \phi_j = \delta_{ij}, \quad i,j=1,\ldots,s-1.
\end{equation}
We recall that these differentials are of the form
\begin{equation}\label{nho}
	\rmd \phi_j(z) = \frac{p_j(z)}{w(z)}\rmd z,
\end{equation}
where $p_j(z)$ are polynomials of degree not greater than $s-2$. Likewise, we denote by $ \rmd \Omega_0$
the third kind normalized meromorphic differential
\begin{equation}
	\label{norddb}	
	\oint_{A_i} \rmd \Omega_0=0, \quad i=1,\ldots,s-1,
\end{equation}
whose only poles are $\infty_1$ and $\infty_2$, and such that
\begin{equation}
	\label{nordd}
	 \rmd \Omega_0(z)
	=
	\left\{\begin{array}{ll}
		(  z^{-1}+\mathcal{O}(z^{-2}) )\rmd z, &\quad z\rightarrow \infty_1,\\
		(-z^{-1}+\mathcal{O}(z^{-2}) )\rmd z, &\quad z\rightarrow \infty_2.
	\end{array}\right.
\end{equation}
It can be written as
\begin{equation}\label{deo}
	\rmd \Omega_0(z) = \frac{P_0(z)}{w(z)}\rmd z,
\end{equation}
where $P_0(z)$ is a polynomial of degree $s-1$.  In appendix~A we derive the following expression
for the second-order derivatives of the prepotential,
\begin{equation}
	\label{ab20}
	\frac{\partial^2 \mathcal{F}}{\partial S_i\partial S_j}
	= 
	4 \pi\rmi (\delta_{js}-1) \phi_j(a_{i}^+) - 2\,\Omega_0 (a_{i}^+),
\end{equation}
where the corresponding Abelian integrals $\phi_j$ and $\Omega_0$ are defined in appendix~A
by integration in the first sheet of the Riemann surface. Therefore, the field equations~(\ref{fie})
admit the following general formulation in terms of Abelian integrals:
\begin{equation}
	\label{sua}
	4\pi\rmi
	\sum_{j=1}^{s-1} N_j\,\phi_j(a_{i}^+) + 2 N \Omega_0 (a_{i}^+)-\log \Lambda^{2N}=0,\quad i=1,\ldots,s.
\end{equation}
%%%%%%%%%%%%%%%%%%%%%%%%%%%%%%%%%%%%%%%%%%%%%%%%%%%%%%%%%%%%%
\subsection{The  case $s=1$}
%%%%%%%%%%%%%%%%%%%%%%%%%%%%%%%%%%%%%%%%%%%%%%%%%%%%%%%%%%%%%
In the one-cut case if we denote $a_1^-=a$, $a_1^+=b$, $S=S_1$ and
$L=L_1$  then, differentiating~(\ref{pre1}) with respect to $S$, we find
\begin{equation}
	\label{eq:ell}
	L = \int_a^b W(z) \frac{\partial\rmd q(z)}{\partial S} + S \frac{\partial^2 \mathcal{F}}{\partial S^2},
\end{equation}
and
\begin{equation}
	\label{ome}
	\rmd \Omega_0(z)=\frac{\rmd z}{\sqrt{(z-a)(z-b)}}.
\end{equation}
Hence~(\ref{ab20}) reduces to~\cite{FE03b,FE03,FE03d}
\begin{equation}\label{eq:sec}
	\frac{\partial^2 \mathcal{F}}{\partial S^2} = -\Log\left(\frac{b-a}{4}\right)^2.
\end{equation}
Using~(\ref{ome}) and~(\ref{dersw}) (cf.~appendix~A)  it follows easily that
\begin{equation}
	\frac{\partial a}{\partial S}=\frac{4}{h(a)\,(a-b)},\quad \frac{\partial b}{\partial S}=\frac{4}{h(b)\,(b-a)},
\end{equation}
and then (\ref{eq:sec}) implies 
\begin{equation}\label{d3}
		\frac{\partial^3 \mathcal{F}}{\partial S^3} = -\frac{8}{(b-a)^2}\,\Big(\frac{1}{h(a)} +\frac{1}{h(b)}\Big),
\end{equation}
where $h(z)$ is given by~(\ref{ache}).

Several useful results follow from these formulas. For instance, by substituting recursively~(\ref{ome})
into~(\ref{eq:ell}) and into~(\ref{pre1}) we derive the following general expression for the prepotential in the one-cut case:
\begin{equation}\label{expf}
	\mathcal{F} = \frac{1}{2} \int_a^b W(z) \rmd q(z)
	                   - \frac{S}{2\pi\rmi} \int_a^b \frac{W(z) \rmd z}{\sqrt{(z-a)(z-b)}}
	                   - \frac{S^2}{2} \Log\left(\frac{b-a}{4}\right)^2.
\end{equation}
Moreover the field equation in the case of unbroken gauge group ($s=1$)
\begin{equation}
	\label{crif}
	N \frac{\partial^2 \mathcal{F}}{\partial S^2}+ \log \Lambda^{2 N}=0,
\end{equation}
takes the form
\begin{equation}
	\label{ssss}
	\log\Big(\frac{\delta}{2 \Lambda}\Big)^{2N}=0,
\end{equation}
where $\delta=(b-a)/2$. Thus there are   $N$ different values of $\delta^2$ characterizing the vacuum states
\begin{equation}
 	\label{dev}
	\delta_k^2 = 4 \zeta_k \Lambda^2,\quad k=1,\ldots,N.
\end{equation}
Furthermore, from~(\ref{d3}) we deduce that the  singular (non-analytic) solutions of the field equation~(\ref{crif})
may only arise  near points $(\mathcal{S}_{\mathrm{c}},\Lambda_{\mathrm{c}})$ such that the one-cut spectral
curve for $S=\mathcal{S}_{\mathrm{c}}$ satisfies one of the conditions
\begin{description}
	\item[(1)] The cut  shrinks to a single point $a=b$. In this case $S_c=0$.
	\item[(2)] A double root of $y^2(z)$ collides with a cut endpoint i.e $h(a)=0$ or $h(b)=0$.
	\item[(3)] It is verified that
\begin{equation}
	\label{critf}
	h(a)=-h(b).
\end{equation}
\end{description}
%%%%%%%%%%%%%%%%%%%%%%%%%%%%%%%%%%%%%%%%%%%%%%%%%%%%%%%%%%%%%
\subsection{Special geometry relations}
%%%%%%%%%%%%%%%%%%%%%%%%%%%%%%%%%%%%%%%%%%%%%%%%%%%%%%%%%%%%%
In this section we show how the special geometry relations follow from our equations~(\ref{eq:dfds})
and~(\ref{perb}), which in turn determine the $L_j$ in terms of $B$-periods. In fact, the special geometry
relations on the spectral curve can be formulated  in several forms depending on the homology
basis used for the Riemann surface~(\ref{spe}) with the two infinities $\infty_1$ and $\infty_2$ removed.
We will apply the scheme of Bilal and Metzger (see section~3.2 of~\cite{BI05})  to the basis
$\{A_i,B_i\}_{i=1}^{s-1}\cup \{\widehat{A},\widehat{B}\}$, where
\begin{equation}
	\label{bm}
 	\widehat{A} = \sum_{j=1}^s A_j,
\end{equation}
and $\widehat{B}$ is a non-compact cycle starting at $\infty_2$ of the second sheet, running to a point
$z_-\in \gamma_s$ and then from $z_+$  to $\infty_1$ on the first sheet.
According to~(\ref{periods}),~(\ref{eq:dfds}) and~(\ref{perb})
\begin{equation}
	\label{sgr0} 
	S_j = -\frac{1}{4 \pi  \rmi } \oint_{A_j}  y(z)  \rmd z ,
	\quad
	\frac{\partial\mathcal{F}}{\partial S_s}- \frac{\partial\mathcal{F}}{\partial S_j}
	=
	\frac{1}{2} \oint_{B_j}\mathrm{y}(z)\rmd z,
\end{equation} 
where the prepotential is considered as a function $\mathcal{F}(\mathbf{S})$ of the partial
{}'t~Hooft parameters $\mathbf{S}=(S_1,\ldots,S_s)$.  If instead we consider the prepotential
as a function $\mathcal{F}(S,\widetilde{\mathbf{S}})$, where $S$ is the total {}'t~Hooft parameter and 
$\widetilde{\mathbf{S}}=(S_1,\ldots,S_{s-1})$, we obtain
\begin{equation}
	\label{sgr00} 
	\widetilde{S}_j = -\frac{1}{4 \pi  \rmi } \oint_{A_j}  y(z)  \rmd z ,
	\quad  \frac{\partial}{\partial \widetilde{S}_j}\mathcal{F}(S,\widetilde{\mathbf{S}})
	=
	-\frac{1}{2} \oint_{B_j}\mathrm{y}(z)\rmd z.
\end{equation} 
Furthermore, it is clear that
\begin{equation}
	\label{rig0}
	S = -\frac{1}{4 \pi  \rmi} \oint_{\widehat{A}}  y(z)  \rmd z,
\end{equation}
while the integral of $\mathrm{y}(z)  \rmd z$ on $\widehat{B}$ is divergent. Therefore we introduce a real cut-off $\Lambda$
and take a cycle $\widehat{B}_{\Lambda}$ starting at  $\Lambda$ of the second sheet, running to point
$z_-\in \gamma_s$ and then from $z_+$ to $\Lambda$ on the first sheet. Using the same procedure
as in the derivation of~(\ref{perb}) it follows that
\begin{equation}
	\label{rig01}
	\oint_{\widehat{B}_{\Lambda}}\mathrm{y}(z)\rmd z=2\left(W(\lambda)-2\,g(\Lambda)-L_s\right).
 \end{equation}
 Hence we get the cut-off independent result
\begin{equation}
	\label{rig1}
 	\frac{\partial}{\partial S}\mathcal{F}(S,\widetilde{\mathbf{S}})
	=
	\lim_{\Lambda\rightarrow \infty}
	\left(-\frac{1}{2} \oint_{\widehat{B}_{\Lambda}}\mathrm{y}(z)\rmd z+W(\lambda)-S\,\log \Lambda^2\right).
\end{equation}
The identities~(\ref{sgr0})--(\ref{rig1}) constitute the special geometry relations with respect to the basis 
$\{A_i,B_i\}_{i=1}^{s-1}\cup \{\widehat{A},\widehat{B}\}$. Finally, note that 
\begin{equation}
	\label{rii}
	L_j = \frac{\partial}{\partial S}\mathcal{F}(S,\widetilde{\mathbf{S}})
	+ (1-\delta_{js})\, \frac{\partial}{\partial \widetilde{S}_j}\mathcal{F}(S,\widetilde{\mathbf{S}}),
\end{equation}
which shows that the special geometry relations allow us to determine the parameters $L_j$ in terms
of $B$-periods of the Riemann surface with the infinities removed.
%%%%%%%%%%%%%%%%%%%%%%%%%%%%%%%%%%%%%%%%%%%%%%%%%%%%%%%%%%%%%
\section{ Phase structure and critical processes \label{sec:crit}}
%%%%%%%%%%%%%%%%%%%%%%%%%%%%%%%%%%%%%%%%%%%%%%%%%%%%%%%%%%%%%
\subsection{Critical spectral curves}
%%%%%%%%%%%%%%%%%%%%%%%%%%%%%%%%%%%%%%%%%%%%%%%%%%%%%%%%%%%%%
In order to formulate the notion of critical spectral curves we first recall that a point of a real curve $F(x,y)=0$
is critical if at that point both partial derivatives $\partial F/\partial x$ and $\partial F/\partial y$ vanish.
In the case of a minimal cut $\gamma_j$ of a spectral curve $\Sigma(\gamma, \mathbf{S})$
we have that $F=\re  G_j$ and
\begin{eqnarray}
	\frac{\partial}{\partial x} \re  G_j(z) & = & \re \left(\rme^{-\rmi\arg S_j} y(z)\right),\\
	\frac{\partial}{\partial y} \re  G_j(z) & = &- \im \left(\rme^{-\rmi\arg S_j} y(z)\right).
\end{eqnarray}
Therefore the cut $\gamma_j$ has a critical point if for a certain value  $\mathbf{S}_\mathrm{c}$
of the set of  't~Hooft parameters  a zero $z_0$ of
$y(z)$ different from $a_j^{\pm}$ meets the path $\gamma_j$. In this case the minimal character of the
cut is lost because the phase of $y(z){\rm d}z$ at $z_0$ is undefined, and we
say that  the corresponding spectral curve $\Sigma(\gamma,\mathbf{S}_\mathrm{c})$ is critical. 
As we will illustrate in the study of the cubic model, instances of these critical spectral curves happen
in phase transition processes of splitting  of minimal cuts.  In general critical spectral curves
are common limits of several families with different number of cuts and they arise when a zero $z_0$ of
$y(z)$ different from $a_j^{\pm}$ meets one of the Stokes lines emerging from $a_j^{\pm}$.

Critical  spectral curves exhibit not only splitting of cuts but also birth and death of  cuts  at a distance
as well as merging of two or more cuts~\cite{BE11}. It should be noticed that the merging of two minimal
cuts with partial 't Hooft parameters  $S_1$ and $S_2$  gives rise to a minimal cut only if $\arg S_1=\arg S_2$. 
%%%%%%%%%%%%%%%%%%%%%%%%%%%%%%%%%%%%%%%%%%%%%%%%%%%%%%%%%%%%%
\subsection{Prepotential and its derivatives at the splitting of a cut}
%%%%%%%%%%%%%%%%%%%%%%%%%%%%%%%%%%%%%%%%%%%%%%%%%%%%%%%%%%%%%
Our main application of the discussion of section~\ref{sec:abe} concerns the behavior of the prepotential and its
derivatives at a splitting  of a cut.
Thus let us  consider a family  of $(s-1)$-cut spectral curves  such that the minimal cut $\gamma_{m-1}$ splits 
into two minimal cuts $\widetilde{\gamma}_{m-1}$ and $\widetilde{\gamma}_m$ with
\begin{equation}
	\label{dos}
	\tilde{a}_{m-1}^{+} = \tilde{a}_m^{-} =\alpha,
\end{equation}
to give a new $s$-cut spectral curve. If we denote by super indices ${}^{(s-1)}$ and ${}^{(s)}$ the respective
magnitudes, it follows at once that
\begin{equation}
	\label{equ}
	w^{(s)}(z) = (z-\alpha)w^{(s-1)}(z),
	\quad
	\rho^{(s)}(z)=\rho^{(s-1)}(z).
\end{equation}
The corresponding critical values ${\bf S}^{(s-1)}$ and ${\bf S}^{(s)}$ of the partial {}'t~Hooft parameters are related by
\begin{equation}
	\label{ssps}
	S_k^{(s-1)}
	=
	\left\{ \begin{array}{ll}
 		S_k^{(s)}, & \mbox{for $1\leq k\leq m-2$},\\
 		S_{m-1}^{(s)}+ S_{m}^{(s)},& \mbox{for $k=m-1$},\\
  		S_{k+1}^{(s)}, & \mbox{for $m\leq  k\leq s-1$},
		\end{array}
	\right.
\end{equation}
where $\arg S_{m-1}^{(s-1)}=\arg S_{m-1}^{(s)}=\arg S_{m}^{(s)}$.
In appendix~B we show that the third kind normalized differential $ \rmd \Omega_0$
and the normalized holomorphic differentials $\rmd  \phi_i$ satisfy
\begin{eqnarray}
	\label{ido}
	\rmd \Omega_0^{(s-1)} & = & \rmd \Omega_0^{(s)},\\
	\label{ido2}
	\rmd \phi_k^{(s-1)} & = &\rmd \phi_{\overline{k}}^{(s)},\quad  \mbox{for $1\leq k\leq s-1$},
	\end{eqnarray}
 where 
\begin{equation}
	\label{kbar}
	\overline{k}
	=
	\left\{ \begin{array}{ll}
 		k, & \mbox{for $1\leq k\leq m-2$},\\
  	\mbox{$m-1$ (or $m$)} & \mbox{for $k=m-1$},\\
  	k+1, & \mbox{for $m\leq  k\leq s-1$}.
\end{array}\right.
\end{equation}

As a consequence of~(\ref{pre}), (\ref{prep}), (\ref{ab20}), (\ref{equ}), (\ref{ido}) and~(\ref{ido2})
we have the following relations for the prepotentials  $\mathcal{F}^{(s-1)}$ and $\mathcal{F}^{(s)}$
and their first and second order derivatives at their respective critical values  ${\bf S}^{(s-1)}$ and ${\bf S}^{(s)}$:
\begin{eqnarray}
	\label{equ1}
	\mathcal{F}^{(s-1)} & = & \mathcal{F}^{(s)},\\
	\label{two2}
	\frac{\partial \mathcal{F}^{(s-1)}}{\partial S_k^{(s-1)}}
	& = &
	\frac{\partial \mathcal{F}^{(s)}}{\partial S_{\overline{k}}^{(s)}},
	\quad \mbox{for $1\leq k\leq s-1$},\\
	\label{two3}
	\frac{\partial^2 \mathcal{F}^{(s-1)}}{\partial S_i^{(s-1)}\partial S_j^{(s-1)}}
	& = &
	\frac{\partial^2 \mathcal{F}^{(s)}}{\partial S_{\overline{i}}^{(s)}\partial S_{\overline{j}}^{(s)}},
	\quad  \mbox{for $1\leq  i,j\leq s-1$}.
\end{eqnarray}

Let us consider now a family of spectral curves parametrized by a real \emph{control parameter} $T$
such that  at a certain critical value $T=T_\mathrm{c}$  there is a splitting of one cut,
i.e for $T<T_\mathrm{c}$  ($T>T_\mathrm{c}$) the spectral curves have one minimal cut
(two minimal cuts). Let us assume that at the critical value $T_\mathrm{c}$  
\begin{equation}
	S_1^{(1)} = S_1^{(2)} + S_2^{(2)},
	\quad
	\dot{S}_1^{(1)} = \dot{S}_1^{(2)} + \dot{S}_2^{(2)},
	\quad
	\ddot{S}_1^{(1)}=\ddot{S}_1^{(2)}+\ddot{S}_2^{(2)},
\end{equation}
where dots stand for derivatives with respect to $T$. Then from  (\ref{equ1})--(\ref{two3}) 
we have that at the critical value $T_\mathrm{c}$
\begin{eqnarray}
	\mathcal{F}^{(2)} & = & \mathcal{F}^{(1)},
	\label{eq:df0}\\
	\frac{\rmd \mathcal{F}^{(2)}}{\rmd T}
	& = &
	\frac{\partial \mathcal{F}^{(2)}}{\partial S_1^{(2)}} \dot{S}_1^{(2)}
	+
	\frac{\partial \mathcal{F}^{(2)}}{\partial S_2^{(2)}} \dot{S}_2^{(2)}
	\nonumber\\
	& = &
	\frac{\rmd \mathcal{F}^{(1)}}{\rmd S_1^{(1)}} (\dot{S}_1^{(2)}+\dot{S}_2^{(2)}) 
	=
	\frac{\rmd \mathcal{F}^{(1)}}{\rmd T},
	\label{eq:df1}\\
	\frac{\rmd^2 \mathcal{F}^{(2)}}{\rmd T^2}	
	& = &
	\frac{\partial \mathcal{F}^{(2)}}{\partial S_1^{(2)}} \ddot{S}_1^{(2)}
	+
	\frac{\partial \mathcal{F}^{(2)}}{\partial S_2^{(2)}}  \ddot{S}_2^{(2)}
	+
	\sum_{i,j=1}^2 \frac{\partial \mathcal{F}^{(2)}}{\partial S_i^{(2)}\partial S_j^{(2)}}
	\dot{S}_i^{(2)} \dot{S}_j^{(2)}
	\nonumber\\
	& = &
	\frac{\rmd \mathcal{F}^{(1)}}{\rmd S_1^{(1)}} (\ddot{S}_1^{(2)}+\ddot{S}_2^{(2)})
	+
	\frac{\rmd^2 \mathcal{F}^{(1)}}{\rmd {(S_1^{(1)})}^2} (\dot{S}_1^{(2)}+\dot{S}_2^{(2)})^2
	=
	\frac{\rmd^2 \mathcal{F}^{(1)}}{\rmd T^2}.
	\label{eq:df2}
\end{eqnarray}	
Thus the prepotential and its first two derivatives are continuous at $S=S_{\mathrm{c}}$.
However, examples in random matrix theory show a jump
discontinuity  for  the third order derivative~\cite{GR80,BL03,AL10}.
Therefore  the splitting  of one cut is expected to be generically a third-order phase transition
in the space of spectral curves with minimal cuts. 
%%%%%%%%%%%%%%%%%%%%%%%%%%%%%%%%%%%%%%%%%%%%%%%%%%%%%%%%%%%%%
\subsection{Splitting of one  cut in the quantum parameter space}\label{sec:vmtc}
%%%%%%%%%%%%%%%%%%%%%%%%%%%%%%%%%%%%%%%%%%%%%%%%%%%%%%%%%%%%%
For simplicity let us consider a critical point $S_1^{(1)}$ of a one-cut family of spectral curves corresponding
to a splitting into two cuts with partial 't Hooft parameters $S_1^{(2)}$ and  $S_2^{(2)}$.
Then from (\ref{two3})  we have that at these critical values 
\begin{equation}
		\label{two3b}
	\frac{\partial^2 \mathcal{F}^{(1)}}{\partial (S_1^{(1)})^2}=
		\frac{\partial^2 \mathcal{F}^{(2)}}{\partial S_{i}^{(2)}\partial S_{j}^{(2)}}.
	\quad  \mbox{for $ i,j=1,2$}.
\end{equation}
As a consequence the sectors $\mathcal{M}_{\rm q}^{(1)}$ and $\mathcal{M}_{\rm q}^{(2)}$
of the quantum space of parameters touch at the value of $\Lambda^2$ given by
\begin{equation}
	\label{lac}
	\Lambda_{\mathrm{c}}^2
	= \exp{\Big(-\frac{\partial^2 \mathcal{F}^{(1)}}{\partial (S_1^{(1)})^2}\Big)}.
\end{equation}
Indeed, it is clear that $S_1^{(1)}$ and $(S_1^{(2)},S_2^{(2)})$ satisfy the field equations
\begin{equation}
	\label{f1}
	N \frac{\partial^2 \mathcal{F}^{(1)}}{\partial S^2}+ \log \Lambda_{\mathrm{c}}^{2 N}=0,
\end{equation}
and 
\begin{equation}
	\label{f2}
	\sum_{i=1}^2 N_i
	\frac{\partial^2 \mathcal{F}^{(2)}}{\partial S_i \partial S_j}+ \log \Lambda_{\mathrm{c}}^{2 N}=0,\quad j=1,2,
\end{equation}
respectively. In this way we have that a  critical spectral curve corresponding to a one-cut splitting
determines an interpolating point  between  the  vacua spaces  corresponding to the phases of
unbroken $U(N)$ and broken $U(N_1)\times U(N_2)$ gauge groups.
%%%%%%%%%%%%%%%%%%%%%%%%%%%%%%%%%%%%%%%%%%%%%%%%%%%%%%%%%%%%
\section{The cubic model in the one-cut case \label{sec:onecc}}	
%%%%%%%%%%%%%%%%%%%%%%%%%%%%%%%%%%%%%%%%%%%%%%%%%%%%%%%%%%%%
In this section we apply our theoretical results of sections~\ref{sec:sc} and~\ref{sec:prep} to provide a fairly
complete global description of the spectral curves with one minimal cut for the cubic potential, which we write
in a form resembling the standard form of the exponent in the integral expression of the Airy function,
\begin{equation}
	\label{eq:w3}
	W(z) = \frac{z^3}{3} - w z.
\end{equation}
%%%%%%%%%%%%%%%%%%%%%%%%%%%%%%%%%%%%%%%%%%%%%%%%%%%%%%%%%%%%
\subsection{Endpoints, series expansions and prepotential}
%%%%%%%%%%%%%%%%%%%%%%%%%%%%%%%%%%%%%%%%%%%%%%%%%%%%%%%%%%%%
We will denote $a_1^{-}= a$, $a_1^{+}= b$, and $S=S_1$. Following the procedure outlined in
section~\ref{sec:dec} the corresponding $y$ function to be substituted in~(\ref{eq:c1}) has the form
\begin{equation}
	\label{eq:w3y1}
	y(z) = \left(z+\frac{1}{2}(a+b)\right) \sqrt{(z-a)(z-b)},
\end{equation}
and $f(z)=-4S z+b_0$. The resulting equations for $a$ and $b$ are simpler when expressed in terms of their
semi-sum $\beta=(b+a)/2$ and semi-difference $\delta=(b-a)/2$,
\begin{equation}
	\label{eq:y1h1}
	2 \beta^2 + \delta^2 = 2 w,
\end{equation}
\begin{equation}
	\label{eq:y1h2}
	\beta \delta^2 = 2 S.
\end{equation}
Therefore $\beta$  satisfies the cubic equation
\begin{equation}
	\label{eq:beta}
	\beta^3 - w \beta + S = 0,
\end{equation}
and $\delta$ is determined by
\begin{eqnarray}
	\label{eq:delta1}
	\delta^2 & = & \frac{2 S}{\beta}, \quad \mbox{if $\beta\neq0$}\\
	\label{eq:delta2}
	\delta^2 & = & 2 w, \quad \mbox{if $\beta=0$}.
\end{eqnarray}
Note that in the latter case $S$ must be zero.

The solutions $\beta(w,S)$ of the cubic equation~(\ref{eq:beta})---as well as $\delta(w,S)$ and the endpoints $a(w,S)$
and $b(w,S)$---satisfy the scaling relation
\begin{equation}
	\label{eq:1cutscaling}
	\beta(w,S) = w^{1/2} \beta(1,S/w^{3/2}),
\end{equation}
so that, aside from the $w^{1/2}$ factor, these magnitudes are functions of the single complex variable
$S/w^{3/2}$. In our context it is natural to fix the value of $w$ (i.e., to fix the potential) so that the solutions
of the endpoint equations for the  one-cut case of the cubic model are described by the three-sheeted
genus zero Riemann surface (\ref{eq:beta}). The corresponding three branches of $\beta$ are given by
\begin{equation}
	\label{eq:betak}
	\beta_k(S) = - \frac{w}{3 \Delta_k(S)} - \Delta_k(S), \quad (k=0,1,2)
\end{equation}
where
\begin{equation}
	\Delta_k(S) = \rme^{\rmi 2\pi k/3}
	                     \sqrt[3]{\frac{S}{2} + \sqrt{\frac{S^2}{4} - \left(\frac{w}{3}\right)^3}}.
\end{equation}
Here we assume that the cubic root has nonnegative real part and that the square root has nonnegative imaginary part.
The finite branch points where two roots coalesce are
\begin{equation}
	S_\pm =  \pm 2 (w/3)^{3/2},
\end{equation}
at which $\beta_0(S_-)=\beta_2(S_-)=-\sqrt{w/3}$ and $\beta_1(S_+)=\beta_2(S_+)=\sqrt{w/3}$.
%%%%%%%%%%%%%%%%%%%%%%%%%%%%%%%%%%%%%%%%%%%%%%%%%%%%%%
\begin{figure}
	\begin{center}
		\includegraphics[width=14cm]{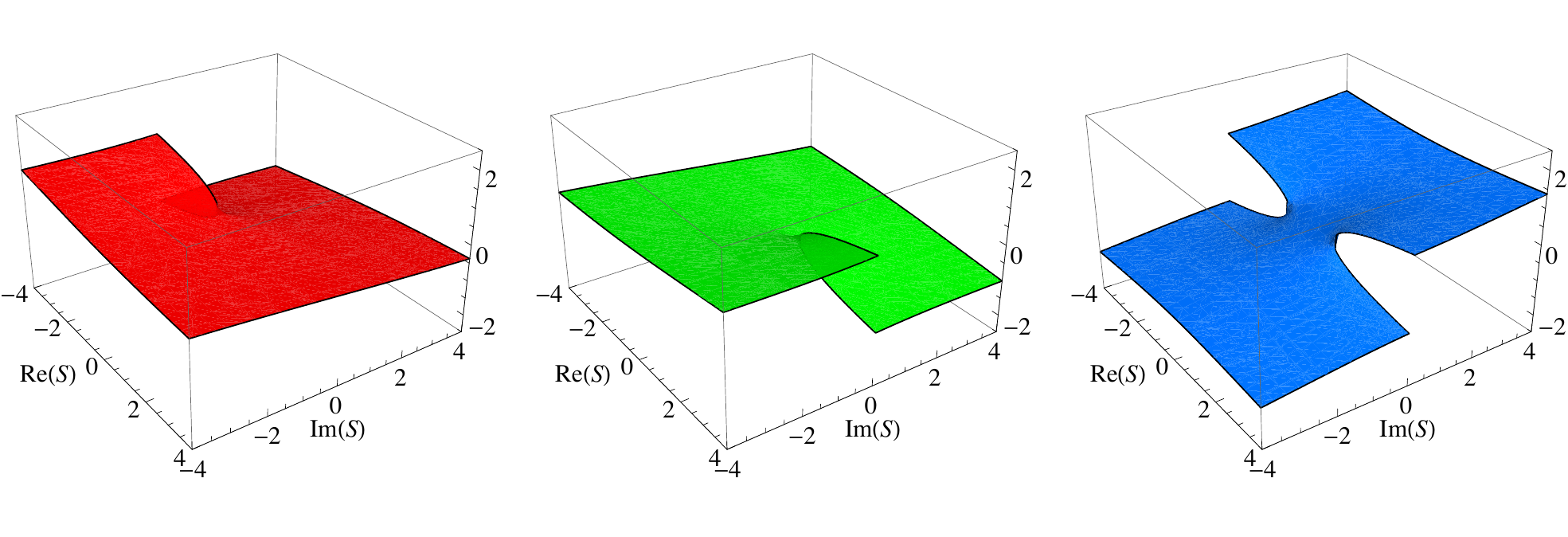}
	\end{center}
	\caption{$\im\beta_0(S)$, $\im\beta_1(S)$,  and $\im\beta_2(S)$ for $w=3/2^{2/3}$.
	              For this value of $w$ the branch points are $S_\pm=\pm 1$.\label{fig:rs}}
\end{figure}
%%%%%%%%%%%%%%%%%%%%%%%%%%%%%%%%%%%%%%%%%%%%%%%%%%%%%%

In the three plots of figure~\ref{fig:rs} we show the imaginary parts $\im\beta_0(S)$, $\im\beta_1(S)$,
and $\im\beta_2(S)$ respectively of the three branches of  $\beta(S)$ for a fixed
value $w=3/2^{2/3}$. This value of $w$ has been chosen because the corresponding branch points
are $S_\pm=\pm1$, and therefore easily identifiable in the figure. The upper and lower edges of the cuts
along $\re S \le S_-=-1$ of $\beta_0(S)$ (red surface) and $\beta_2(S)$ (blue surface) have to be glued,
as well as the edges of the cuts along $\re S \ge S_+=1$ of $\beta_1(S)$ (green surface)
and $\beta_2(S)$ (blue surface).

Thus, given a value of $S\neq S_\pm$, there are three values $\beta_k(S)$ from which we find three
possible pairs of endpoints,
\begin{equation}
	\label{eq:akws}
	a_k(S) = \beta_k(S) - \sqrt{2 S/\beta_k(S)},
\end{equation}
\begin{equation}
	\label{eq:bkws}
	b_k(S) = \beta_k(S) + \sqrt{2 S/\beta_k(S)}.
\end{equation}
 Using~(\ref{eq:betak}) we find the 
expansions of $\beta_k(S)$ as $S\rightarrow 0$
\begin{eqnarray}
	\label{eq:b0s}
	\beta_0(S) & = & - \sqrt{w} - \frac{S}{2w} + \frac{3 S^2}{8 w^{5/2}} + \mathcal{O}(S^3),\\
	\label{eq:b1s}
	\beta_1(S) & = &   \sqrt{w} - \frac{S}{2w} - \frac{3 S^2}{8 w^{5/2}} + \mathcal{O}(S^3),\\
	\label{eq:b2s}
	\beta_2(S) & = & \frac{S}{w}+\mathcal{O}(S^3).
\end{eqnarray}
Therefore according to our discussion in section~\ref{sec:sccl}, as $S\rightarrow 0$  the branches
$\beta_0$ and $\beta_1$ represent families of one-cut spectral curves with a classical limit  in which
the cuts shrink to the  critical point $-\sqrt{w}$ and $+\sqrt{w}$ of the cubic potential.
Note also that the endpoints $a$ and $b$ can be expanded as Puiseux series. For example,
\begin{equation}
	\label{eq:1cutapms}
	a_1^{\pm}(S) =  \sqrt{w}
                                      \pm \frac{2^{1/2} S^{1/2}}{w^{1/4}}
	                              - \frac{S}{2w}
                                      \pm \frac{S^{3/2}}{2^{3/2}w^{7/4}}
	                              - \frac{3 S^2}{8 w^{5/2}}
	                              + \mathcal{O}(S^{5/2}),
\end{equation}
where $a_1^-\equiv a$ and $a_1^+\equiv b$. The function $\beta_2(S)$ does not represents a family of
spectral curves with a shrinking minimal cut as $S\rightarrow 0$ since it reduces to the solution~(\ref{eq:delta2})
with  $a_2^{\pm}\rightarrow \pm \sqrt{2w}$. We will see in sections~\ref{sec:cc} and \ref{sec:sli} that there are
not spectral curves with a minimal cut corresponding to $\beta_2(S)$ near $S= 0$ and that the limit
$S\rightarrow 0$ represents a critical  spectral curve with two minimal cuts such that $S_1+S_2=0$.
 
The expression for the prepotential~(\ref{expf}) takes, up to an unessential constant term $-w^3/12$, 
the following simple form in terms of the function $\beta(S)$
\begin{equation}\label{pbs}
	\mathcal{F} = \frac{\beta^6}{2} 
	                   - \frac{5}{12} w \beta^4 - \frac{(w \beta-\beta^3)^2}{2} \Log\left(\frac{w-\beta^2}{2}\right).
\end{equation}
This expression for the prepotential is equivalent to previous results in which the first
two terms in the right-hand side appear in rational form. For example, equation~(51) in~\cite{BR78}
can be put in our polynomial form using~equation~(49) of~\cite{BR78} to eliminate the denominator. 
%%%%%%%%%%%%%%%%%%%%%%%%%%%%%%%%%%%%%%%%%%%%%%%%%%%%%%%%%%%%%
\subsection{Superpotentials and the quantum parameter space}
%%%%%%%%%%%%%%%%%%%%%%%%%%%%%%%%%%%%%%%%%%%%%%%%%%%%%%%%%%%%%
The one-cut case can be used to determine the vacua structure of the unbroken gauge group $U(N)$~\cite{FE03d}.
Using~(\ref{pbs}) and taking into account that
\begin{equation}
	\frac{\partial \beta}{\partial S}=\frac{1}{w-3\beta^2},
\end{equation}
the superpotential can be expressed in terms of $\beta$ as
\begin{eqnarray}
	W_{{\rm eff}}
	& = & -\frac{2}{3} N \beta^3-N(w\beta-\beta^3)
	           \Log\left(\frac{w-\beta^2}{2}\right)
	           + (w\beta-\beta^3)\log\Lambda^{2N} \nonumber\\
	& = & -\frac{2}{3} N \beta^3
	         + (w\beta-\beta^3)\log\Big(\frac{2\,\Lambda^2}{w-\beta^2}\Big)^N. \label{wb}
\end{eqnarray}

We can obtain the vevs $\mathcal{S}$ since~(\ref{dev}),~(\ref{eq:y1h1}) and~(\ref{eq:y1h2}) imply
\begin{equation}\label{vacc2}
	(\delta^{(k)})^2 = 4\, \Lambda^2 \zeta_k,\quad \beta^{(k,\pm)} =\pm\, \sqrt{w-2\Lambda^2\zeta_k},\quad  k=1,\ldots,N,
\end{equation}
and 
\begin{equation}
	\label{vacc}
	\mathcal{S}^{(k,\pm)} = \pm\, 2\Lambda^2 \zeta_k \sqrt{w-2 \Lambda^2\zeta_k},\quad k=1,\ldots,N.
\end{equation}
Hence  we find two families of $N$  quantum vacua $|k,\pm\rangle$ which correspond to the two
classical vacua $\pm \sqrt{w}$ of the cubic model.

To calculate the low-energy superpotential $W_{{\rm low}}(w,\Lambda)$ associated with the $|k,\pm\rangle$
vacua we notice that from (\ref{vacc2}) it follows that
\begin{equation}
	\frac{2 \Lambda^2 \zeta_k}{w-(\beta^{(k,\pm)})^2}=1,
\end{equation}
so that~(\ref{wb}) leads to the following simple exact expression
\begin{equation}
 	\label{lowe}
 	W_{{\rm low}}^{(k,\pm)}
	=
	-\frac{2}{3} N (\beta^{(k,\pm)})^3
	=
	\mp\frac{2}{3} N (w-2 \Lambda^2 \zeta_k)^{3/2}.
\end{equation}
We can now calculate the vevs
 \begin{equation}
	\langle\Tr \Phi\rangle^{(k,\pm)}
	=
	-\frac{\partial W_{{\rm low}}^{(k,\pm)} }{\partial w}
	=
	\pm N \sqrt{w-2\Lambda^2\zeta_k}
	=
	N\beta^{(k,\pm)},
\end{equation}
and check that
\begin{equation}
	\mathcal{S}^{(k,\pm)}
	=
	\frac{\partial W_{{\rm low}}^{(k,\pm)} }{\partial \log \Lambda^{2N}}
	=
	2 \Lambda^2 \beta^{(k,\pm)} \zeta_k.
\end{equation}

In this way we have that  the one-cut sector  $\mathcal{M}_{\rm q}^{(1)}$ of the quantum parameter space
can be decomposed   into $N$ subsets
 \begin{equation}
	\mathcal{M}_{\rm q}^{(1)}=\cup_{k=1}^N \mathcal{M}_{\rm q}^{(1,k)},
 \end{equation}
where $\mathcal{M}_{\rm q}^{(1,k)}$ is represented by the two-sheeted  Riemann surface of the function
$\sqrt{w-\lambda}$.  Here  
\begin{equation}
	\label{sla}
	\lambda= 2\Lambda^2\zeta_k.
\end{equation}
The two sheets $\mathcal{M}_{\rm q}^{(1,k,\pm)}$ are connected through  the branch cuts emerging
from $\lambda=w$. Each point $(\lambda ,\pm)\in \mathcal{M}_{\rm q}^{(1,k,\pm)}$ determines a unique
spectral curve given by
\begin{equation}
	\label{qsc}
	S = \pm\lambda \sqrt{w-\lambda},\quad
	\beta = \pm\sqrt{w-\lambda},\quad \delta^2=2\lambda.
 \end{equation}
%%%%%%%%%%%%%%%%%%%%%%%%%%%%%%%%%%%%%%%%%%%%%%%%%%%%%%%%%%%%
\subsection{Critical spectral  curves in the complex $S$ plane\label{sec:cc}}
%%%%%%%%%%%%%%%%%%%%%%%%%%%%%%%%%%%%%%%%%%%%%%%%%%%%%%%%%%%%
In this section we derive an analytic condition  met by critical curves in the complex $S$ plane, i.e., by
the locus of values of the complex {}'t~Hooft parameter $S$ that yield critical spectral curves with
minimal cuts. Let us denote by $\Sigma_k(S)$ the families of one-cut
spectral curves corresponding to the branches $\beta_k(S)$ for  $k=0,1,2$. The minimal cuts of $\Sigma_k(S)$
for a fixed value of $S$ are Stokes lines defined by 
\begin{equation}
	\label{eq:1cst}
	\re G_k(z) = 0
\end{equation}
where
\begin{equation}
	\label{eq:1cgk}
	G_k(z)
	=
	\rme^{-\rmi\arg S}
	\int_{a_k^{-}}^{z'} \left(z'+\beta_k\right) \sqrt{(z'-\beta_k)^2-\delta_k^2}\,\rmd  z'.
\end{equation}
Therefore, if for a certain value of  $S$ the double zero $-\beta_k$ of $y(z)$ meets the Stokes line, i.e., if 
\begin{equation}
	\label{eq:1ccc}
	\re G_k(-\beta_k) = 0,
\end{equation}
the curve has a critical point. In fact, we can find an analytic condition (which, however, has
to be solved numerically) for this critical behavior, because the integrals~(\ref{eq:1cgk})
can be evaluated in closed form. Thus, we find that the critical curve in the complex
$S$-plane corresponding to a branch $\beta_k(S)$ is given by
\begin{equation}
	\label{eq:1cccex}
	\re\left[
	\frac{w}{3 S}\sqrt{6 \beta_k^2-2w}
	+ \log\left(\sqrt{\frac{\beta_k}{2S}}\left(-2\beta_k+\sqrt{6 \beta_k^2-2w}\right)\right)
	\right]
	= 0.
\end{equation}
%%%%%%%%%%%%%%%%%%%%%%%%%%%%%%%%%%%%%%%%%%%%%%%%%%%%%%%%%%%%%%%
\begin{figure}
	\begin{center}
		\includegraphics[width=8cm]{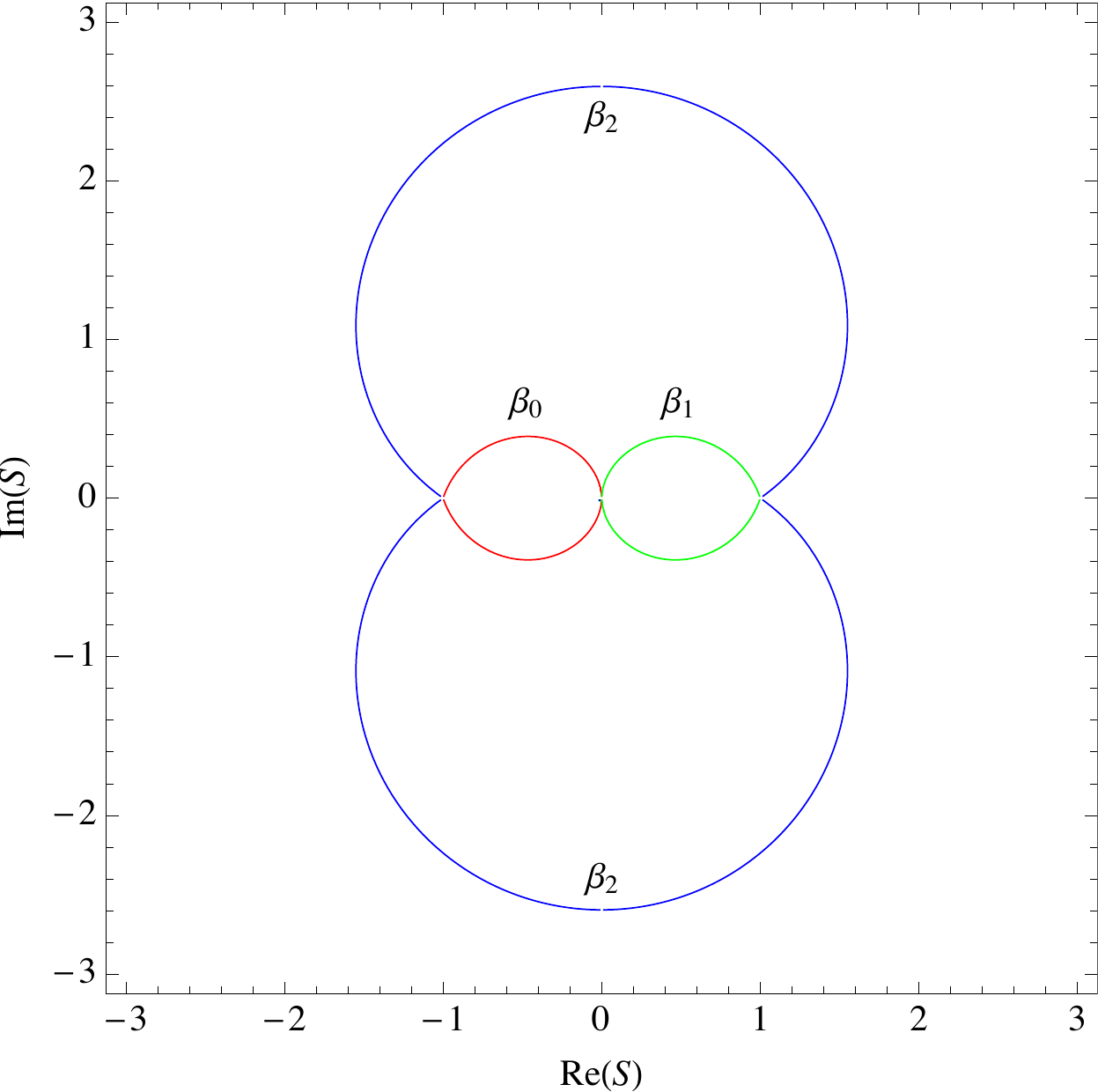}
	\end{center}
	\caption{Critical curves in the sense that $\re G_k(-\beta(S))=0$ for $w=3/2^{2/3}$.
	              Each arc is marked with the corresponding branch $\beta_k$ and drawn
	              in the color matching the corresponding branch in figure~\ref{fig:rs}.\label{fig:cc}}
\end{figure}
%%%%%%%%%%%%%%%%%%%%%%%%%%%%%%%%%%%%%%%%%%%%%%%%%%%%%%%%%%%%%%%

In figure~\ref{fig:cc} we show the three branches  $\re G_k(-\beta)=0\, (k=0,1,2)$ obtained by numerical solution
of~(\ref{eq:1cccex}) for  $w=3/2^{2/3}$. Each arc is marked and colored to match the corresponding branch $\beta_k$
in figure~\ref{fig:rs} (e.g., the red critical curve, marked $\beta_0$, lies on the first plot of figure~\ref{fig:rs}).
We have performed extensive numerical calculations of the corresponding Stokes graphs,
from which we infer the following picture.

\begin{figure}
	\begin{center}
		\includegraphics[width=14cm]{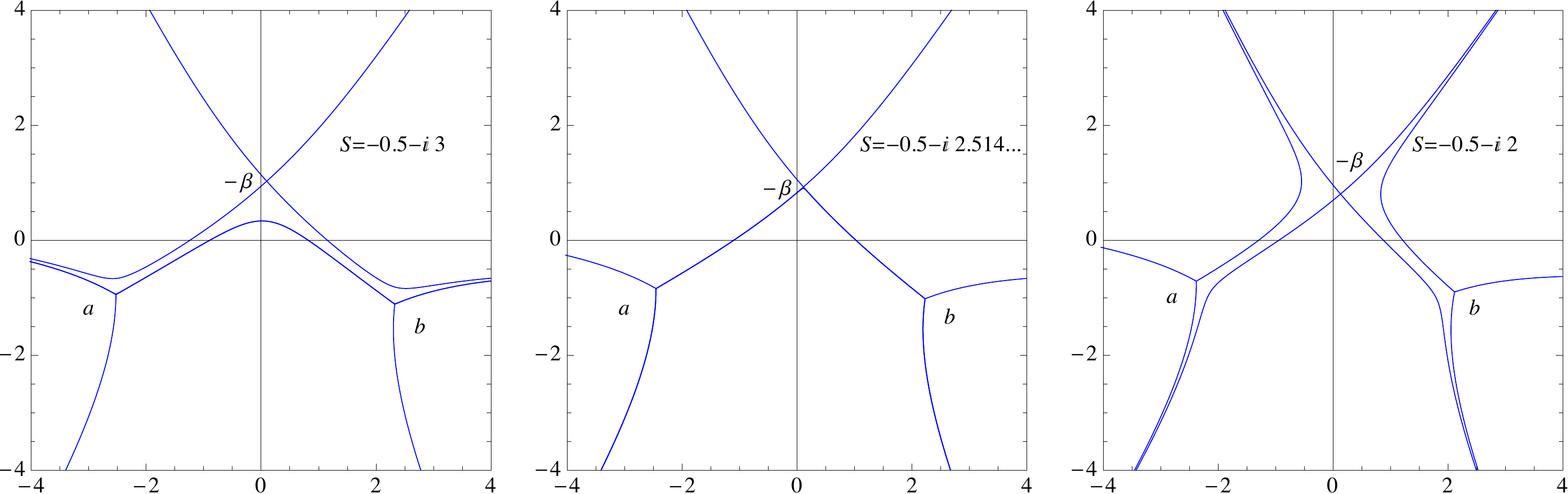}
	\end{center}
	\caption{Stokes complexes on the branch $\beta_2(S)$ for $w=3/2^{2/3}$ and three
	              values of $S$: $S=-0.5-\rmi\,3$ (slightly below the blue critical curve in figure~\ref{fig:cc}),
	              $S=-0.5-\rmi\,2.514\,668\ldots$ (on the critical curve), and
	              $S=-0.5-\rmi\,2$ (slightly above the critical curve).\label{fig:sg2}}
\end{figure}

Consider first the branch $\beta_2(S)$. The critical (blue) curve in figure~\ref{fig:cc} separates the
$S$ plane into a bounded region and an unbounded region. In the unbounded region we always find
a minimal cut joining $a$ to $b$, and therefore a one-cut solution. The critical (blue) line corresponds
to configurations where the double root $-\beta_2$ meets the minimal cut, and in the bounded region
of the $S$ plane the Stokes graphs do not feature finite Stokes lines, i.e., there are not one-cut spectral
curves with a minimal cut in this branch.
This behavior is illustrated in figure~\ref{fig:sg2}, where we show the Stokes graphs for $w=3/2^{2/3}$
and three values of $S$: $S=-0.5-\rmi\,3$, in the unbounded region of the $S$ plane and slightly below the
blue critical curve, $S=-0.5-\rmi\,2.514\,668\ldots$ on the critical curve, and
$S=-0.5-\rmi\,2$ in the bounded region of the $S$ plane but slightly above the critical curve.
Numerical calculations to be presented in the next section show that these critical configurations
can be continued to spectral curves with two minimal cuts. Thus, crossing the blue critical curve
from the bounded to the unbounded region would correspond to a ``merging of two minimal cuts''
in the terminology of random matrix theory.
%%%%%%%%%%%%%%%%%%%%%%%%%%%%%%%%%%%%%%%%%%%%%%%%%%%%%%%%%%%%%%
%%%%%%%%%%%%%%%%%%%%%%%%%%%%%%%%%%%%%%%%%%%%%%%%%%%%%%%%%%%%%

\begin{figure}
	\begin{center}
		\includegraphics[width=14cm]{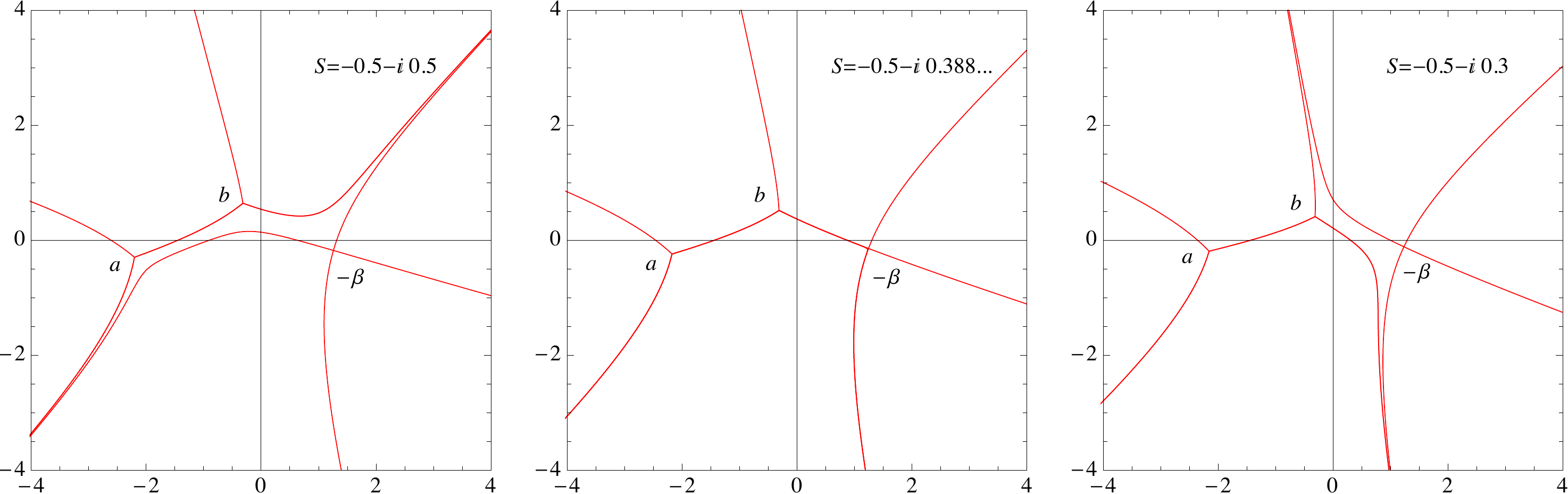}
	\end{center}
	\caption{Stokes complexes on the branch $\beta_0(S)$ for $w=3/2^{2/3}$ and three
	              values of $S$: $S=-0.5-\rmi\,0.5$ (slightly below the red critical curve in figure~\ref{fig:cc}),
	              $S=-0.5-\rmi\,0.388\,126\ldots$ (on the critical curve), and
	              $S=-0.5-\rmi\,0.3$ (slightly above the critical curve).\label{fig:sg0}}
\end{figure}

The behavior of the branches $\beta_0(S)$ and $\beta_1(S)$ is the same, and qualitatively different from
the behavior of $\beta_2(S)$. For concreteness we will describe the behavior of $\beta_0(S)$.
The corresponding critical curve (the red curve in figure~\ref{fig:cc})
separates the $S$ plane into a bounded region and an unbounded region. In the unbounded region we always find
a minimal cut joining $a$ to $b$. The critical curve corresponds
to configurations where the double root $-\beta_0$ does not meet the minimal cut joining $a$ to $b$,
but a second finite Stokes line joining $b$ to $-\beta_0$ appears. However, if we proceed to the bounded region
of the $S$ plane the Stokes graphs again feature a finite Stokes line joining $a$ to $b$, i.e., there are
one-cut solutions with a minimal cut. This behavior is illustrated in figure~\ref{fig:sg0},
where we show the Stokes graphs for $w=3/2^{2/3}$
and three values of $S$:  $S=-0.5-\rmi\,0.5$, in the unbounded region of the $S$ plane and slightly below the
red critical curve,  $S=-0.5-\rmi\,0.388\,126\ldots$ on the critical curve, and
$S=-0.5-\rmi\,0.3$ in the bounded region of the $S$ plane but slightly above the critical curve.
Numerical calculations to be presented in the next section also show that crossing the critical curve
of the $S$ plane in the $\beta_0$ branch from the unbounded to the bounded region we can continue 
with one-cut solutions with a minimal cut (as illustrated in figure~\ref{fig:sg0}) or alternatively
we can continue  from the critical configuration to a spectrum curve with two minimal cuts.
In this sense, crossing the red critical curve from the bounded to the unbounded region may correspond either
to no phase change or to a ``birth of a minimal cut at a distance'' in the terminology of random
matrix theory~\cite{EY06}.
%%%%%%%%%%%%%%%%%%%%%%%%%%%%%%%%%%%%%%%%%%%%%%%%%%%%%%%%%%%%
%%%%%%%%%%%%%%%%%%%%%%%%%%%%%%%%%%%%%%%%%%%%%%%%%%%%%%%%%%%%
\subsection{Spectral curves with one minimal cut in the quantum parameter space}
\begin{figure}
	\begin{center}
		\includegraphics[width=10cm]{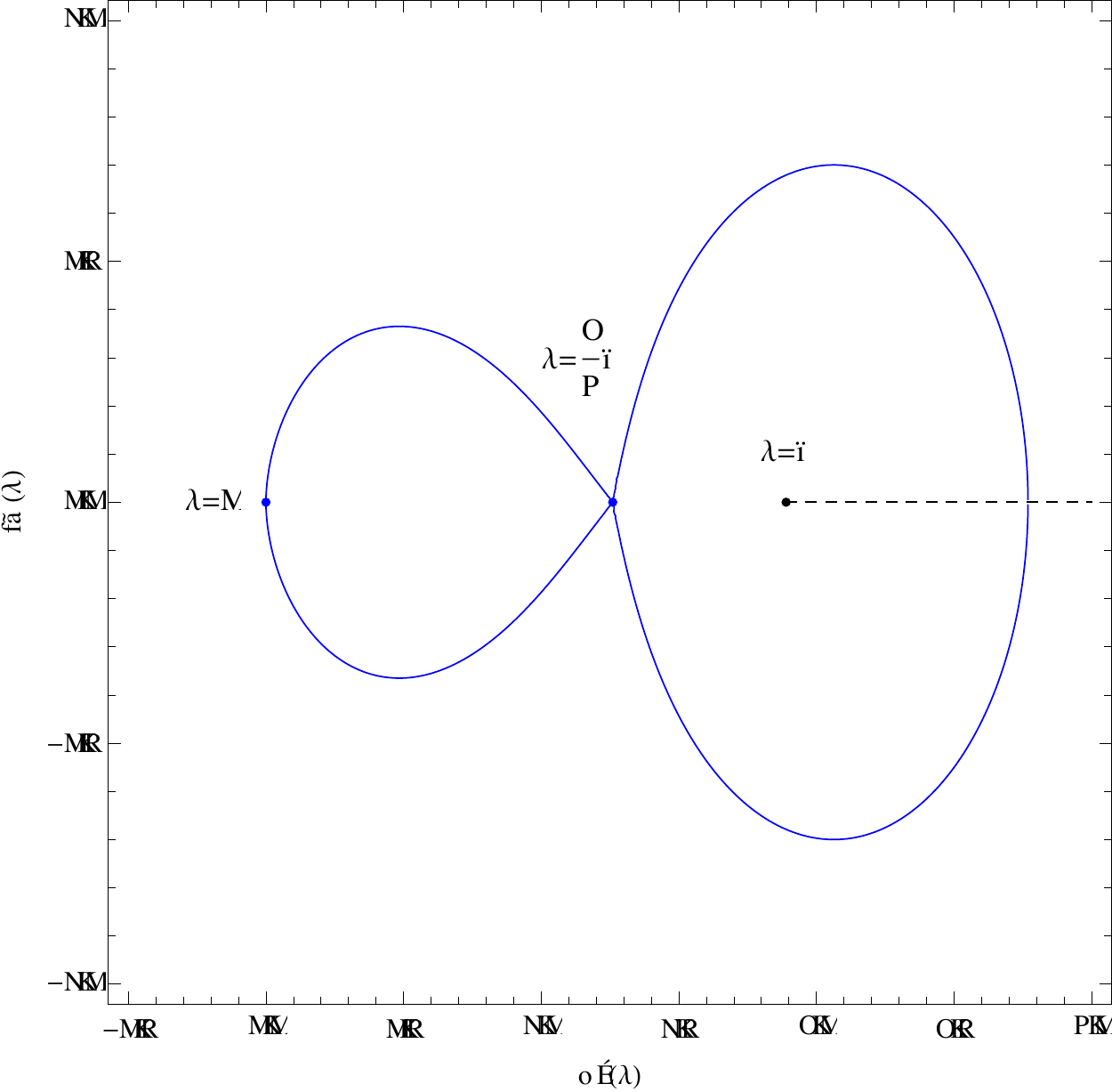}
	\end{center}
	\caption{Critical curve in $\mathcal{M}_{\rm q}^{(1,k,\pm)}$ for $w=3/2^{2/3}$.\label{fig:cclambda}}
\end{figure}

%%%%%%%%%%%%%%%%%%%%%%%%%%%%%%%%%%%%%%%%%%%%%%%%%%%%%%%%%%%%
In the case of the cubic model  each  point $(\lambda ,\pm)\in \mathcal{M}_{\rm q}^{(1,k,\pm)}$
determines a spectral curve~(\ref{qsc}) which will be critical if it satisfies the condition~(\ref{eq:1cccex}) or,
equivalently, in terms of the variable $\lambda$
\begin{equation}
	\label{eq:1cccex2b}
	\re\left[\pm
	\frac{w}{3\lambda} \frac{\sqrt{4w-6\lambda}}{\sqrt{w-\lambda}}
	+ \log\left(
	\frac{\sqrt{4w-6\lambda}\mp2\sqrt{w-\lambda}}{\sqrt{2\lambda}}
	\right)
	\right]
	= 0.
\end{equation}
This equation defines a curve in $\mathcal{M}_{\rm q}^{(1,k,\pm)}$ and  since
\begin{equation}
	\left(\frac{\sqrt{4w-6\lambda}-2\sqrt{w-\lambda}}{\sqrt{2\lambda}}\right)
	\left(\frac{\sqrt{4w-6\lambda}+2\sqrt{w-\lambda}}{\sqrt{2\lambda}}\right)
	= -1,
\end{equation}
the parts of the curve lying in each sheet $\mathcal{M}_{\rm q}^{(1,k,+)}$ and $\mathcal{M}_{\rm q}^{(1,k,-)}$
are identical. Figure~\ref{fig:cclambda} shows that this critical curve is composed of two ovals, one of them
containing the branch point $\lambda=w$. 
%%%%%%%%%%%%%%%%%%%%%%%%%%%%%%%%%%%%%%%%%%%%%%%%%%%%%%%%%%%%
%%%%%%%%%%%%%%%%%%%%%%%%%%%%%%%%%%%%%%%%%%%%%%%%%%%%%%%%%%%%

It is not difficult to give an argument that suggests which regions of the sheets $\mathcal{M}_{\rm q}^{(1,k,\pm)}$
correspond to spectral curves with one minimal cut:
\begin{enumerate}
	\item As $\lambda\to0$ then $S(\lambda)\to 0$ and $\beta(\lambda)\to\sqrt{w}$. Therefore,
	         according to~(\ref{eq:b1s}),
	         the function $\beta(\lambda)$ near $\lambda = 0$ behaves as $\beta_1(S)$ near $S=0$.
	\item As $\lambda\to w$ then $S(\lambda)\to 0$ and $\beta(\lambda)\to 0$. Therefore,
	         according to~(\ref{eq:b2s}),
	         the function $\beta(\lambda)$ near $\lambda = w$ behaves as $\beta_2(S)$ near $S=0$.
\end{enumerate}
As a consequence, the analysis in sec.~\ref{sec:cc} permits us to conjecture that the points in the interior of the
oval containing $\lambda=w$ do not supply  spectral curves with one minimal cut,
and that the oval itself describes processes of splitting of a cut. Conversely, 
points in the interior of the leftmost oval determine spectral curves with one minimal cut and points
on the oval lead to critical spectral curves describing birth of a cut at a distance.
Finally, since as $\lambda\to \infty$ then $S(\lambda)\to \infty$ so that according to the analysis in
sec.~\ref{sec:cc} we expect that points outside the critical ovals determine spectral curves with one minimal cut.
%%%%%%%%%%%%%%%%%%%%%%%%%%%%%%%%%%%%%%%%%%%%%%%%%%%%%%%%%%%%
\begin{figure}
	\begin{center}
		\includegraphics[width=14cm]{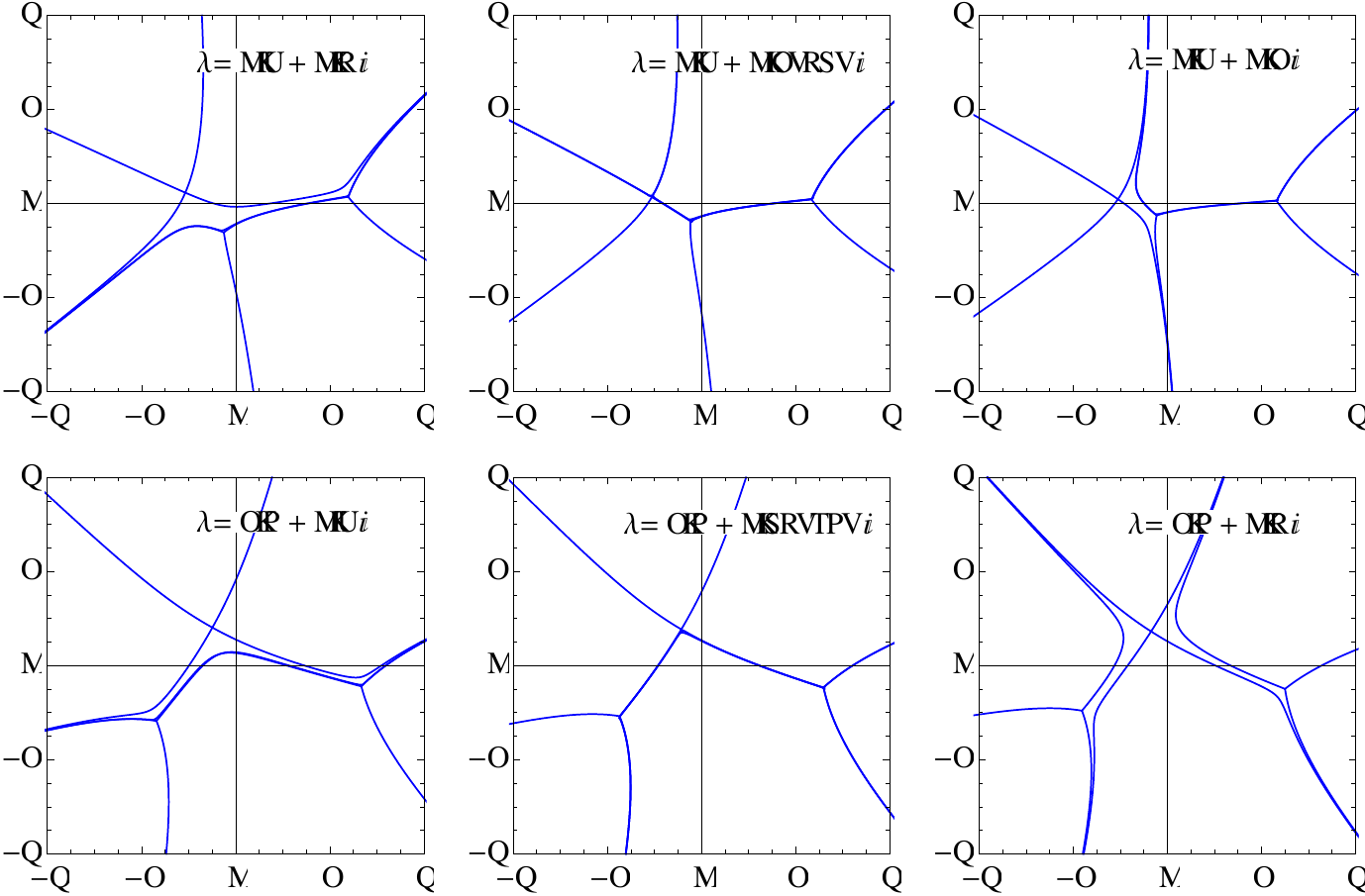}
	\end{center}
	\caption{Stokes complexes for two paths in $\mathcal{M}_{\rm q}^{(1,k,\pm)}$
	              which enter vertically from above the left oval (first row)
	              and the right oval (second row) of the critical
	              curve of figure~\ref{fig:cclambda}.\label{fig:pathslambda}}
\end{figure}
%%%%%%%%%%%%%%%%%%%%%%%%%%%%%%%%%%%%%%%%%%%%%%%%%%%%%%%%%%%%
These arguments are supported by numerical calculations, two of which are illustrated in
figure~\ref{fig:pathslambda}. The three graphs in the first row correspond to three values of $\lambda$
on the same vertical $\re\lambda=0.8$. The first value is above the left oval, the second value is critical,
i.e., on this oval, and the third value  is already in the interior of the oval. These graphs show how we proceed
from a spectral curve with one minimal cut, through a critical configuration (in which the three turning points are 
joined by finite Stokes lines), to another spectral curve with one minimal cut.  The three graphs in the second
row correspond to three values of $\lambda$ on the vertical $\re\lambda=2.3$. The first value is above the right
oval, the second value is critical, i.e., on this oval, and the third value is already in the interior of the oval. Now
the graphs show that we proceed from a spectral curve with one minimal cut, through a critical configuration
(in which the three turning points are  joined by finite Stokes lines), but that as we enter the interior of the
oval it does not exist an spectral curve with one minimal cut.

As we discussed in section~\ref{sec:vmtc}  critical spectral curves describing a splitting  of a cut process determine
interpolating points between the sectors  $\mathcal{M}_{\rm q}^{(1)}$ and $\mathcal{M}_{\rm q}^{(2)}$ of the quantum
parameter space. Therefore for the cubic model the ovals containing $\lambda=w$  in the sheets
$\mathcal{M}_{\rm q}^{(1,k,\pm)}$ are interpolating curves between the subsets of these sectors corresponding
to spectral curves with minimal cuts.  These results show a structure of the quantum parameter space  drastically different
from that found in~\cite{FE03d}  where,  due to the absence of a concrete choice of the cuts,
the only interpolation found  between the sectors $\mathcal{M}_{\rm q}^{(1)}$ and $\mathcal{M}_{\rm q}^{(2)}$
is the single point $\lambda=w$.  However, as it will be proved in section~\ref{sec:sli},
this point describes a merging of two minimal cuts into a non minimal cut.
%%%%%%%%%%%%%%%%%%%%%%%%%%%%%%%%%%%%%%%%%%%%%%%%%%%%%%%%%%%%
\section{Two-cut spectral curves in the cubic model\label{sec:twocc}}
%%%%%%%%%%%%%%%%%%%%%%%%%%%%%%%%%%%%%%%%%%%%%%%%%%%%%%%%%%%%
In this section we consider the two-cut phase of spectral curves for the cubic potential~(\ref{eq:w3}).
We first derive the series expansions around the critical points of the potential  for the endpoints of two-cut
spectral curves with classical limit. Then we discuss the particular solution corresponding to the 
slice $S_1+S_2=0$. Finally, we describe a numerical method to calculate spectral curves with
two minimal cuts and their critical processes.
%%%%%%%%%%%%%%%%%%%%%%%%%%%%%%%%%%%%%%%%%%%%%%%%%%%%%%%%%%%%
\subsection{Endpoints and series expansions}
%%%%%%%%%%%%%%%%%%%%%%%%%%%%%%%%%%%%%%%%%%%%%%%%%%%%%%%%%%%%
In the two-cut case we will denote $a_1^{-}\equiv a$, $a_1^{+}\equiv b$, $a_2^{-}\equiv c$ and
$a_2^{+}\equiv d$. Following again the procedure outlined in section~\ref{sec:dec} the corresponding
$y$ function to be substituted in~(\ref{eq:c1}) has the form
\begin{equation}
	\label{eq:w3y2}
	y(z) = \sqrt{(z-a)(z-b)(z-c)(z-d)},
\end{equation}
and $f(z)=-4(S_1+S_2) z+b_0$. The equations for the possible endpoints turn out to be
\begin{eqnarray}
	\label{eq:y2h1}
	a b c + a b d + a c d + b c d & = & 4 (S_1 + S_2), \\
	\label{eq:y2h2}
	a b + a c + a d+ b c  + b d + c d & = & - 2 w, \\
	\label{eq:y2h3}
	a + b + c + d & = & 0, \\
	\label{eq:y2h4}
	I(a,b,c,d) & = & 2\pi\rmi S_1,
\end{eqnarray}
where
\begin{equation}
	\label{eq:iabcd}
 	I(a,b,c,d) = \int_a^b \sqrt{(z_+-a)(z_+-b)(z_+-c)(z_+-d)}\,\rmd z.
\end{equation}

Although~(\ref{eq:iabcd}) can be expressed in terms of elliptic integrals it is clear that, except
in specially simple particular cases, the system~(\ref{eq:y2h1})--(\ref{eq:y2h4}) cannot be solved
in closed form. However, we can characterize solutions that have a classical limit by means of
their series expansions for the endpoints  that tend to the critical points $\pm\sqrt{w}$ of the cubic
potential as $S_1$ and $S_2$ tend to zero. These series are the analog of
equation~(\ref{eq:1cutapms}) in the one-cut case. Since in the two-cut case
we lack closed-form solutions analog to~(\ref{eq:betak}), we resort to direct substitution
of suitable Puiseux expansion into equations~(\ref{eq:y2h1})--(\ref{eq:y2h4}). We do have,
however, an scaling relation analog to~(\ref{eq:1cutscaling}):
\begin{equation}
	a(w,S_1,S_2) = w^{1/2} a(1,S_1/w^{3/2},S_2/w^{3/2}),
\end{equation}
with similar relations for $b$, $c$ and $d$. Therefore we can write the following Puiseux
expansions:
\begin{eqnarray}
	\label{eq:puiseuxa}
	a & = & \sqrt{w} \left(-1 + \sum_{j=1}^\infty \frac{a_j}{w^{3j/4}}\right),\\
	\label{eq:puiseuxb}
	b & = & \sqrt{w} \left(-1 + \sum_{j=1}^\infty \frac{(-1)^j a_j}{w^{3j/4}}\right),\\
	\label{eq:puiseuxc}
	c & = & \sqrt{w} \left(1 + \sum_{j=1}^\infty \frac{c_j}{w^{3j/4}}\right),\\
	\label{eq:puiseuxd}
	d & = & \sqrt{w} \left(1 + \sum_{j=1}^\infty \frac{(-1)^j c_j}{w^{3j/4}}\right),	
\end{eqnarray}
where we have separated explicitly the first term to identify $a$ and $b$ as the endpoints
of the cut that opens up from $-\sqrt{w}$ and $c$ and $d$ as the endpoints of the cut that
opens up from $\sqrt{w}$, and where the coefficients $a_j$ and $c_j$ are functions
of $S_1$ and $S_2$. Substitution of these series into the first three
equations~(\ref{eq:y2h1})--(\ref{eq:y2h3}) is straightforward. As to the fourth equation~(\ref{eq:y2h4}),
which involves the integral~(\ref{eq:iabcd}), we can formally fix the endpoints of the contracting
integration interval with a linear change of variable
\begin{equation}
	I(a,b,c,d) = -\rmi(b-a)^2 \sqrt{(c-a)(d-a)}\int_0^1 \sqrt{t(1-t)} r_{ab}(t)\rmd t,
\end{equation}
where
\begin{equation}
	\label{eq:rab}	
   	r_{ab}(t) = \sqrt{\left(1-\frac{b-a}{c-a}t\right)\left(1-\frac{b-a}{d-a}t\right)}.
\end{equation}
Then we expand $r_{ab}(t)$ as a power series in $t$ around $t=0$ (recall that $b-a$ tends to zero
while the denominators remain finite) and integrate term by term. All the integrals are of the form
\begin{equation}
	\label{eq:inttype1}
	\int_0^1 \sqrt{t(1-t)} t^k\,\rmd t
	=
	\frac{\sqrt{\pi}}{2}
	\frac{\Gamma(k+3/2)}{\Gamma(k+3)},
	\quad
	k=0,1,2,\ldots
\end{equation}
and the substitution of the Puiseux series for $a$, $b$, $c$, and $d$ in the resulting expression
is again straightforward.

To illustrate the pattern of the resulting Puiseux series we show the first six coefficients $a_k$ and $c_k$:
\begin{eqnarray}
	a_1& = & \rmi\sqrt{2 S_1},\\
	a_2& = & \frac{1}{2}(-{S_1}+{S_2}),\\
	a_3& = & -\rmi\frac{\sqrt{2S_1}}{8} (2 {S_1}-3 {S_2}),\\
	a_4& = & \frac{1}{8}(3 S_1^2-8 {S_1} {S_2}+3 S_2^2),\\
	a_5& = & \rmi\frac{\sqrt{2S_1}}{128}(36 S_1^2-122 S_1 S_2+69 S_2^2),\\
	a_6& = & -\frac{1}{32}({S_1}-{S_2})(16 S_1^2-59 S_1 S_2+16 S_2^2),\\
	c_1& = & -\sqrt{2 S_2},\\
	c_2& = & \frac{1}{2}({S_1}-{S_2}),\\
	c_3& = &  \frac{\sqrt{2S_2}}{8} (3 {S_1}-2 {S_2}),\\
	c_4& = & -\frac{1}{8}(3 S_1^2-8 S_1 S_2+3 S_2^2)\\
	c_5& = & -\frac{\sqrt{2S_2}}{128}(69 S_1^2-122 S_1 S_2+36 S_2^2),\\
	c_6& = & \frac{1}{32}({S_1}-{S_2})(16 S_1^2-59 S_1 S_2+16 S_2^2).
\end{eqnarray}
The odd coefficients $a_k$ and $c_k$ involve the square roots of the corresponding {}'t~Hooft
parameters (and cancel in the series for the differences $b-a$ and $d-c$), while the even
coefficients are homogeneous polynomials of degree $k/2$ in $S_1$ and $S_2$.
Note also that if in the expressions for the $a_k$ we set $S_1=0$ and $S_2=S$ we formally
recover the one-cut expansions~(\ref{eq:1cutapms}).
%%%%%%%%%%%%%%%%%%%%%%%%%%%%%%%%%%%%%%%%%%%%%%%%%%%%%%%%%%%%%%
\subsection{The cubic model on the slice $S_1+S_2=0$\label{sec:sli}}
%%%%%%%%%%%%%%%%%%%%%%%%%%%%%%%%%%%%%%%%%%%%%%%%%%%%%%%%%%%%%%
As an explicit application of the endpoint equations~(\ref{eq:y2h1})--(\ref{eq:y2h4}) we describe now
the two-cut spectral curves of the cubic model in the Seiberg-Witten slice $S_1+S_2=0$~\cite{SE94}.
These configurations have been used in the literature to study the phase structure of brane/anti-brane systems
at large $N$ and non-perturbative effects in matrix models~\cite{CA07,HE07,HE08,KL10}.   
We look for solutions of~(\ref{eq:y2h1})--(\ref{eq:y2h4}) satisfying
\begin{equation}
	\label{syms}
	c = -b,
	\quad
	d = -a,
\end{equation}
where we assume $|a|>|b|$. Then these equations reduce to
\begin{equation}
	\label{first}
	a^2 + b^2 = 2 w,
\end{equation}
\begin{equation}
	\label{second}
	\int_{a}^b \sqrt{(z_+^2-a^2)(z_+^2-b^2)} \rmd  z = 2 \pi\rmi S_1.
\end{equation}
We may solve~(\ref{first}) in the form
\begin{equation}
	\label{set1}
	a = |a| \rme^{\rmi\arg w/2},
	\quad
	b = |b| \rme^{\rmi\arg w/2},
	\quad
	|b| = \sqrt{2 |w| - |a|^2},
\end{equation}
and it is immediate to see that  the straight line segments $[a, b]$ and $[c,d]$ are minimal cuts provided that
\begin{equation}
	\label{sli}
	\arg S_1 =\frac{3}{2} \arg w.
\end{equation}
In particular~(\ref{second}) reads
\begin{equation}
	\label{ins1}
	\int_{\sqrt{2|w|-|a|^2}}^{|a|} \sqrt{(x^2-|a|^2)(x^2+|a|^2-2|w|)} \rmd  x
	=
	2 \pi\rmi |S_1|,
\end{equation}
which gives  $|S_1|$ as a function of $|a|$. 

Therefore~(\ref{sli}) determines a family of spectral curves $\Sigma(\gamma,S_1,S_2=-S_1)$ with two minimal cuts,
parameterized by  the values of $|a|$ in the interval $[\sqrt{|w|},\sqrt{2|w|}]$.  The limit 
$|a|\rightarrow \sqrt{2 |w|}$  of this family deserves a particular attention since the spectral curve
reduces to the  one-cut solution~(\ref{eq:b2s}) determined by        
\begin{equation}
	S=0,\quad \beta=0,\quad \delta^2=2 w.
\end{equation}
However  due to the fact that  $\arg S_1\neq \arg S_2$ the cut resulting from the merging is not a  minimal cut.
This spectral curve corresponds to the singular point $\lambda=w$ of the quantum parameter space
$\mathcal{M}_{\rm q}^{(1)}$ at which the solutions $\mathcal{S}^{(k)}(\Lambda^2)$ of the field equation
are not analytic, as it should be expected since $h(z)=z+\beta$ and the condition~(\ref{critf}) is satisfied.
%%%%%%%%%%%%%%%%%%%%%%%%%%%%%%%%%%%%%%%%%%%%%%%%%%%%%%%%%%%%
\subsection{Numerical calculation of minimal cuts and critical processes}
%%%%%%%%%%%%%%%%%%%%%%%%%%%%%%%%%%%%%%%%%%%%%%%%%%%%%%%%%%%%
A direct numerical solution of the system~(\ref{eq:y2h1})--(\ref{eq:y2h4}) without a suitable, well identified
initial point would be extremely difficult (note that for a fixed value of $w$ the endpoints are functions
of the two complex variables $S_1$ and $S_2$). However, we can take advantage of our knowledge
of the critical curve~(\ref{eq:1cccex}) and the corresponding explicit solutions for the one-cut endpoints
given by~(\ref{eq:betak}), (\ref{eq:akws}) and~(\ref{eq:bkws}), and proceed iteratively by small increments
in $S_1$ and $S_2$ to calculate the solutions of~(\ref{eq:y2h1})--(\ref{eq:y2h4}) at any desired pair
of values $(S_1,S_2)$ using as initial approximation at each step the results of the previous one.

\begin{figure}
	\begin{center}
		\includegraphics[width=14cm]{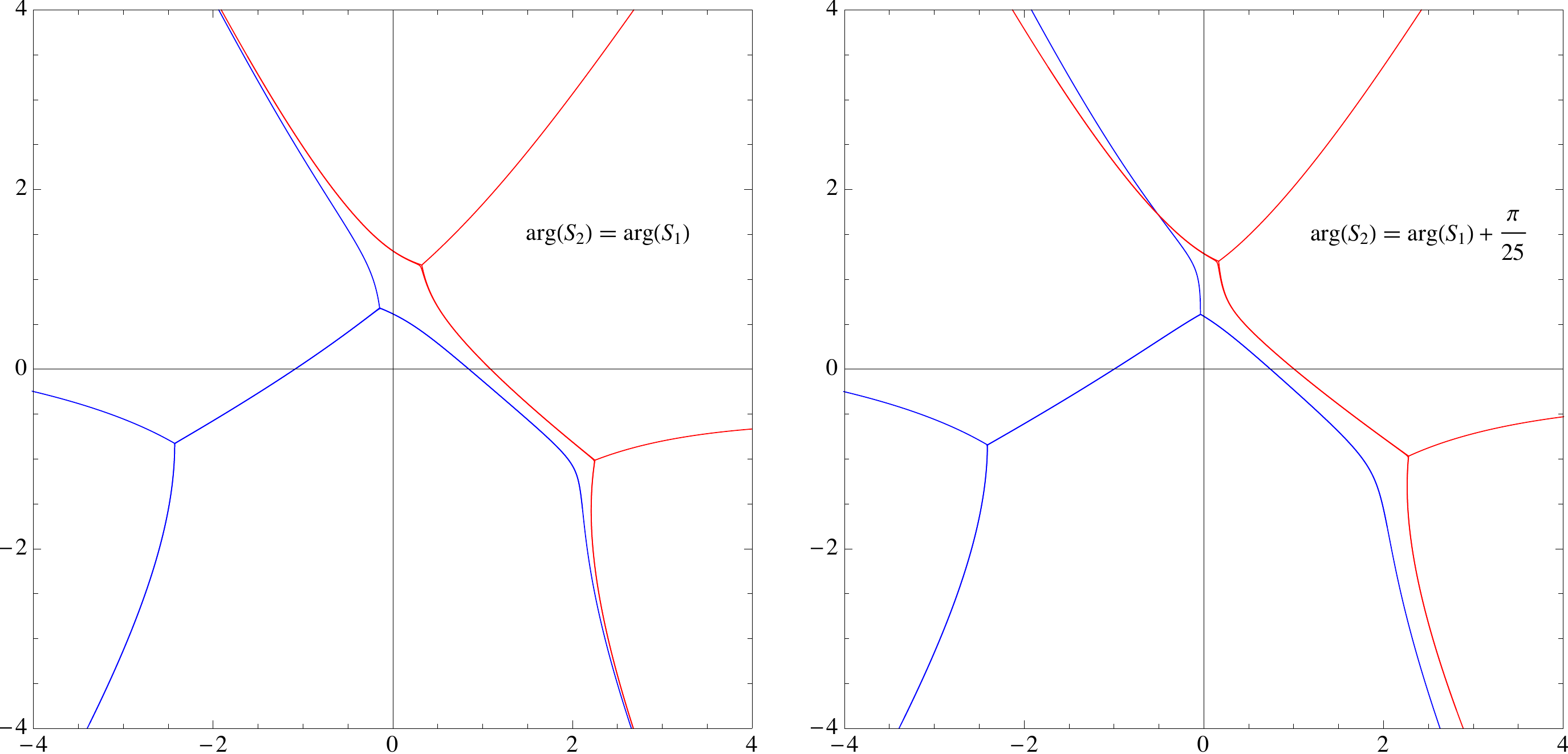}
	\end{center}
	\caption{Stokes complexes in the two-cut region for $w=3/2^{2/3}$ and two pairs $(S_1, S_2)$
	              close to the critical curve: the first graph corresponds to $S_1 = S_2 = 99 S/200$,
	              where $S=-0.5-\rmi\,2.514\,668\ldots$ is the critical value in figure~\ref{fig:sg2}
	              (note that in this case $\arg(S_1)=\arg(S_2)$ and the Stokes lines do not cross);
	              the second graph corresponds to $S_1 = 99 S/200$ and $S_2 =\rme^{\rmi \pi/25} S_1$
	              (note that in this case $\arg(S_1)\neq\arg(S_2)$ and two Stokes lines cross).
	              \label{fig:2c1}}
\end{figure}

To illustrate this approach, consider the second, critical configuration in figure~\ref{fig:sg2}.
In figure~\ref{fig:2c1} we show the numerical continuation from this critical configuration
into the two-cut region for two examples. In the first graph of figure~\ref{fig:2c1} we have set
$S_1 = S_2 = 99 S/200$, where $S=-0.5-\rmi\,2.514\,668\ldots$ is the critical value in
figure~\ref{fig:sg2}. Note that $S_1+S_2$ is a small perturbation of $S$ (cf.~equation~(\ref{ssps})),
i.e., we split the double zero $-\beta$ into two simple zeros of $y(z)^2$ and the critical Stokes graph into a
a graph with two finite Stokes lines of similar length (recall that the critical configuration is not symmetric)
with $\arg(S_1)=\arg(S_2)$. Since the corresponding functions $G_j(z)$ in~(\ref{mesn})
have the same values of $\arg S_j$, the Stokes lines cannot cross. Similarly,
in the second graph of figure~\ref{fig:2c1} we have perturbed slightly this solution to
$S_1 = 99 S/200$ and $S_2 =\rme^{\rmi \pi/25} S_1$. Now $\arg S_1\neq \arg S_2$
and two of the Stokes lines corresponding to different values of $j$ do cross. As we anticipated
in the previous section, looking at this process backwards we have an instance of a
``merging of two minimal cuts'' in the terminology of random matrix theory.
%%%%%%%%%%%%%%%%%%%%%%%%%%%%%%%%%%%%%%%%%%%%%%%%%%%%%%%%%%%%
%%%%%%%%%%%%%%%%%%%%%%%%%%%%%%%%%%%%%%%%%%%%%%%%%%%%%%%%%%%%

\begin{figure}
	\begin{center}
		\includegraphics[width=14cm]{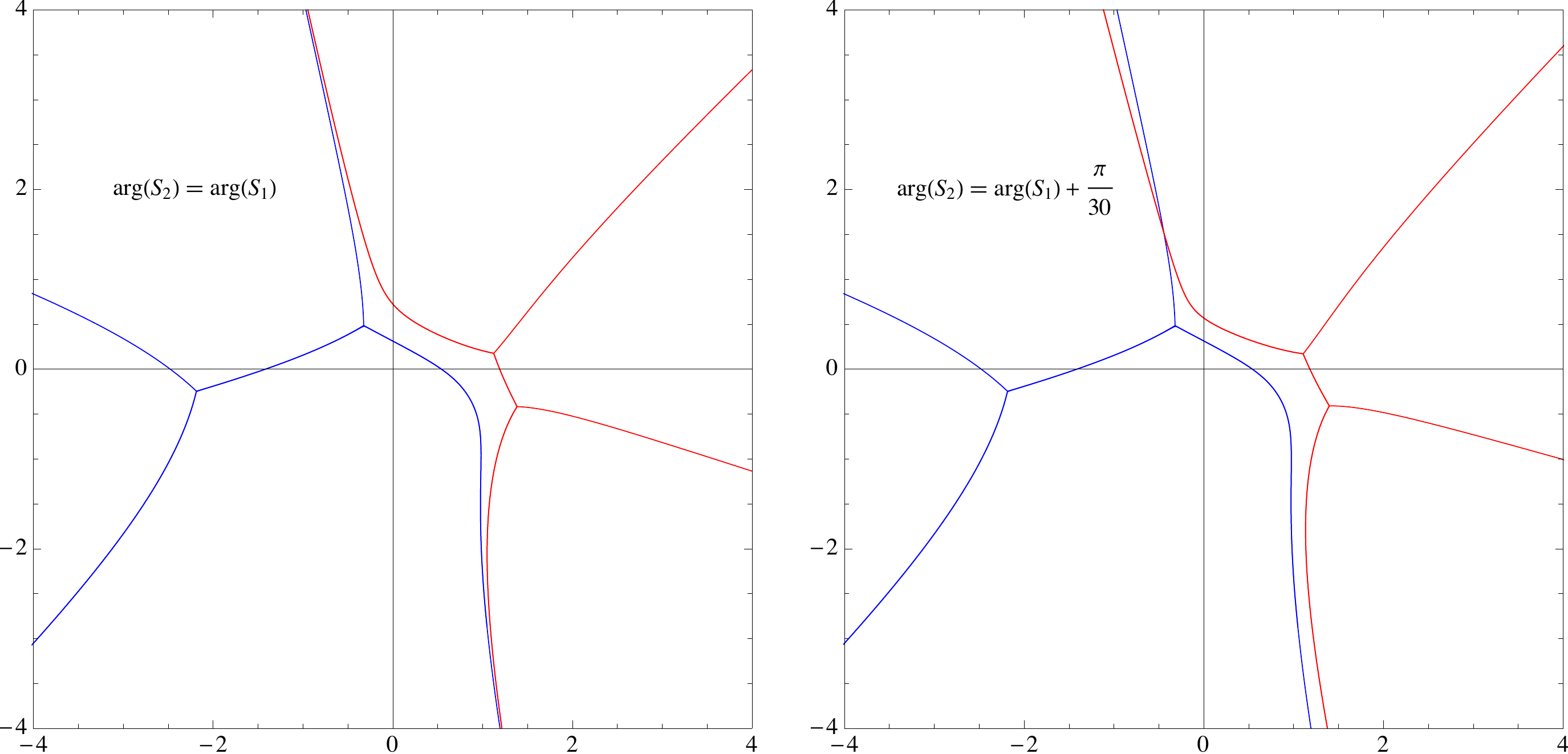}
	\end{center}
	\caption{Stokes complexes in the two-cut region for $w=3/2^{2/3}$ and two pairs $(S_1, S_2)$
	              close to the critical curve: the first graph corresponds to $S_1 = S$, $S_2 = S/10$,
	              where $S=-0.5-\rmi\,0.388\,126\ldots$ is the critical value in figure~\ref{fig:sg0};
	              (note that in this case $\arg(S_1)=\arg(S_2)$ and the Stokes lines do not cross);
	              the second graph corresponds to $S_1 = S$ and $S_2 =\rme^{\rmi \pi/30} S/10$
	              (note that in this case $\arg(S_1)\neq\arg(S_2)$ and two Stokes lines cross).
	              \label{fig:2c2}}
\end{figure}

Likewise, in figure~\ref{fig:2c2} we illustrate the numerical continuation from the critical
configuration in the second graph of figure~\ref{fig:sg0} to the two-cut region.
In the first graph of figure~\ref{fig:2c2} we have set $S_1 = S$, $S_2 = S/10$,
where $S=-0.5-\rmi\,0.388\,126\ldots$ is the critical value in figure~\ref{fig:sg0}.
Again, this is a small perturbation of the critical configuration which we use to generate the
initial approximation. The double zero $-\beta$ splits into two simple zeros and the critical
Stokes graph into a graph with a long finite Stokes line corresponding to $S_1$ and a short
finite Stokes line corresponding to $S_2$. Since in this first graph we have taken
$\arg(S_1)=\arg(S_2)$, the corresponding functions $G_j(z)$ in~(\ref{mesn}) have the same
values of $\arg S_j$, and the Stokes lines cannot cross. In the second graph we have
increased the argument of $S_2$ by $\pi/30$, and two Stokes lines corresponding to different values
of $j$ cross.  As we anticipated in the previous section this is an instance of a
``birth of a minimal cut at a distance'' in the terminology of random matrix theory.
%%%%%%%%%%%%%%%%%%%%%%%%%%%%%%%%%%%%%%%%%%%%%%%%%%%%%%%%%%%%
%%%%%%%%%%%%%%%%%%%%%%%%%%%%%%%%%%%%%%%%%%%%%%%%%%%%%%%%%%%

Therefore we have an efficient numerical algorithm that taking critical one-cut solutions as
data to generate an initial approximation to solve equations~(\ref{eq:y2h1})--(\ref{eq:y2h4}),
allows us to proceed stepwise tracking the solution along any specified path of the complex
variables $S_1$ and $S_2$, generate the two-cut endpoints
$a(S_1,S_2)$, $b(S_1,S_2)$, $c(S_1,S_2)$, $d(S_1,S_2)$ and calculate the corresponding
Stokes graph.
%%%%%%%%%%%%%%%%%%%%%%%%%%%%%%%%%%%%%%%%%%%%%%%%%%%%%%%%%%%%%%
%%  CONCLUSIONS  %%%%%%%%%%%%%%%%%%%%%%%%%%%%%%%%%%%%%%%%%%%%%%%%%%
%%%%%%%%%%%%%%%%%%%%%%%%%%%%%%%%%%%%%%%%%%%%%%%%%%%%%%%%%%%%%%
\section{Summary\label{sec:theend}}
%%%%%%%%%%%%%%%%%%%%%%%%%%%%%%%%%%%%%%%%%%%%%%%%%%%%%%%%%%%%%%
In this paper we have presented a self-contained approach to study the phase transitions
that arise in the study of large $N$ dualities between $\mathcal{N}=1$ SUSY $U(N)$ gauge
theories and string models on local Calabi-Yau manifolds. Conceptually, it is based on the use
of spectral curves with branch cuts that are projections of minimal supersymmetric
cycles. From the practical point of view, the key elements of
our approach are a system of equations for the branch points and a characterization of the
branch cuts as Stokes lines of a suitable set of polynomials. The system of equations
for the branch points is derived from the form of the equation of the spectral curve with a fixed
number of cuts after imposing period relations on the cuts. In turn, these branch cuts are a natural generalization
of the cuts that define the eigenvalue support in holomorphic matrix models (note that in this
latter case the partial {}'t~Hooft parameters are essentially the eigenvalue fractions and
therefore real and positive).  However, we do not rely on any
underlying matrix model but use a variational characterization of the (in general complex) density as
an extremal of a prepotential naturally associated to the spectral curve. 
By writing the derivatives of the prepotential with respect to the {}'t~Hooft
parameters in terms of Abelian integrals we study the splitting of a cut in the space of spectral
cuts and show it is typically a third order transition. We also show that a critical spectral curve
corresponding to the splitting of a cut represents an interpolating points between different
sectors in the quantum parameter space. We have applied these theoretical results and
numerical calculations to study the cubic model, finding the analytic condition satisfied
by critical one-cut spectral curves, and characterizing the transition curves between
the one-cut and two-cut phases both in the space of spectral curves and in the quantum parameter space.
Although our analytic results in the one-cut phase of the cubic model rely on the explicit solution
of a cubic equation, in principle nothing prevents the use of the same numerical methods used
in the two-cut phase to study spectral curves for non polynomial potentials.
%%%%%%%%%%%%%%%%%%%%%%%%%%%%%%%%%%%%%%%%%%%%%%%%%%%%%%%%%%%%
%%  ACK  %%%%%%%%%%%%%%%%%%%%%%%%%%%%%%%%%%%%%%%%%%%%%%%%%%%%%%
%%%%%%%%%%%%%%%%%%%%%%%%%%%%%%%%%%%%%%%%%%%%%%%%%%%%%%%%%%%%
\section*{Acknowledgments}
The financial support of the Universidad Complutense under project GR58/08-910556 and the
Ministerio de Ciencia e Innovaci\'on under projects FIS2008-00200 and FIS2011-22566 are gratefully
acknowledged. 
%%%%%%%%%%%%%%%%%%%%%%%%%%%%%%%%%%%%%%%%%%%%%%%%%%%%%%%%%%%%
%%  APPENDIX  A %%%%%%%%%%%%%%%%%%%%%%%%%%%%%%%%%%%%%%%%%%%%%%%%%%
%%%%%%%%%%%%%%%%%%%%%%%%%%%%%%%%%%%%%%%%%%%%%%%%%%%%%%%%%%%%
\section*{A Prepotential identities}
%%%%%%%%%%%%%%%%%%%%%%%%%%%%%%%%%%%%%%%%%%%%%%%%%%%%%%%%%%%%
In this appendix we collect the proofs of the main identities satisfied by the prepotential associated to
spectral curves. To prove~(\ref{prep}) and~(\ref{iij0}) we differentiate~(\ref{pre}) with respect $S_i$
taking into account that  $\rho(a_1^{\pm})=\cdots=\rho(a_s^{\pm})=0$. Thus we have
\begin{eqnarray}
	\nonumber
	\frac{\partial \mathcal{F}}{\partial S_i}
	& = & 
	\int_{\gamma} W(z) \frac{\partial \rmd q(z)}{\partial S_i}
	-
	\int_{\gamma}\int_{\gamma} \Log(z-z')^2 \rmd q(z')\frac{\partial\rmd q(z)}{\partial S_i}\\
	\nonumber
	& = &
	\int_{\gamma} \left( W(z) - ( g(z_+)+g(z_-) ) \right) \frac{\partial  \rmd q(z)}{\partial S_i}\\
	& = &
	\sum_{j=1}^s L_j \int_{\gamma_j} \frac{\partial  \rmd q(z)}{\partial S_i} =L_i,
\end{eqnarray}
where we have used~(\ref{s1}) and the fact that because of~(\ref{eq:den})
\begin{equation}
	\int_{\gamma_j} \frac{\partial  \rmd q(z)}{\partial S_i}
	=
	\frac{\partial }{\partial S_i}\int_{\gamma_j}  \rmd q(z)=\delta_{ij}.
\end{equation}
Moreover, from~(\ref{s1}) we have that
\begin{equation}
	\label{li1}
	L_j= W(z_j)-\int_{\gamma}  \Log(z-z_j)^2\,\rmd q(z),\quad  z_j\in\gamma_j,
\end{equation}
and~(\ref{prep}) and~(\ref{iij0}) follow immediately.

We will next prove that the asymptotics of $L_j$ in the classical limit is 
\begin{equation}
	\label{elesa}
	L_j \sim W(a_j)-\sum_{k\neq j}S_k\,\log\Delta_{jk}^2+S_j\Big(1+\log\Big(\frac{W''(a_j)}{S_j}\Big)\Big),
\end{equation}
where $\Delta_{jk}\equiv a_j-a_k$. To this aim we set $z_j=a_j^{+}$ in (\ref{li1}) and get
\begin{equation}
	L_j=W(a_j^{+})-\frac{1}{2\,\pi\,i}\int_{\gamma}{\rm Log}(z-a_j^{+})^2\,y(z_+) \rmd z.
\end{equation}
The classical limit of the first term is
\begin{equation}
	\label{waj}
	W(a_j^{+})
	=
	W(\beta_j+\delta_j)\sim W(a_j+\delta_j)\sim W(a_j)+\frac{W''(a_j)}{2} \delta_j^2.
\end{equation}
The second term decomposes as a sum
\begin{equation}
	\label{intl}
	\int_{\gamma}{\rm Log}(z-a_j^{+})^2 y(z_+) \rmd z
	=\sum_{k}\int_{\gamma_k}{\rm Log}(z-a_j^{+})^2 y(z_+) \rmd z.
\end{equation}
Taking into account the periods~(\ref{periods}), for $k\neq j$ we get
\begin{equation}
	\label{intlk}
	\int_{\gamma_k}{\rm Log}(z-a_j^{+})^2 y(z_+) \rmd z
	\sim 
	\Log(\Delta_{jk}^2) \int_{\gamma_k}y(z_+) \rmd z
	=
        2\pi\rmi S_k \Log(\Delta_{jk}^2),\quad k\neq j.
\end{equation}
Moreover, using~(\ref{34}) we obtain
\begin{equation}
	\int_{\gamma_j}\Log(z-a_j^{+})^2 y(z_+)\rmd z
	\sim
	\rmi W''(a_j) \int_{a_j^{-}}^{a_j^{+}} \Log(z-a_j^{+})^2\sqrt{(z-a_j^{-})(a_j^{+}-z)}\rmd z.
\end{equation}
The integral in the right-hand side can be exactly calculated using the change of variables
$z=a_j^{-}+(a_j^{+}-a_j^{-})\, t$:
\begin{eqnarray}
& & \int_{a_j^{-}}^{a_j^{+}} {\rm Log}(z-a_j^{+})^2 \sqrt{(z-a_j^{-})(a_j^{+}-z)}\rmd z \nonumber\\
& & = 4 \delta_j^2\Log(4 \delta_j^2) \int_0^1 \sqrt{t(1-t)}\rmd t +
         4 \delta_j^2 \int_0^1 \Log(t-1)^2\sqrt{t(1-t)}\rmd t\nonumber\\
& & = \pi \frac{\delta_j^2}{2}\Log(4 \delta_j^2)+\pi \frac{\delta_j^2}{2}(1-\Log 16)\nonumber\\
& & = \pi\frac{\delta_j^2}{2}\Big(1+\Log\frac{\delta_j^2}{4}\Big)
\end{eqnarray}
Hence using~(\ref{epc}) the asymptotics~(\ref{elesa}) follows.

Let us prove now the expression~(\ref{ab20}) of the second derivatives of the prepotential
in terms of Abelian integrals. Differentiating~(\ref{li1}) we find that
\begin{equation}
	\label{iija}
	\frac{\partial^2 \mathcal{F}}{\partial S_i\partial S_j}
	=
	\frac{\partial L_i}{\partial S_j}
	=
	-\int_{\gamma}  \Log(z-z_i)^2 \frac{\partial  \rmd q(z)}{\partial S_j},
	\quad z_i\in\gamma_i,
\end{equation}
where the integrals are independent of the choice of $z_i$ in $\gamma_i$.
From~(\ref{y}) and~(\ref{periods}) it follows that
\begin{equation}
	\label{dersw}
	\frac{\partial}{\partial S_j} \mathrm{y}(z)\rmd z
	=
	-4 \pi\rmi (1-\delta_{js})  \rmd \phi_j-2  \rmd \Omega_0.
\end{equation}
Hence
\begin{equation}
	\label{ab1}
	\frac{\partial  \rmd q(z)}{\partial S_j}
	=
	-2 (1-\delta_{js})\rmd \phi_j(z_+)-\frac{1}{\pi\rmi}  \rmd \Omega_0(z_+).
\end{equation}
Substituting~(\ref{ab1}) into~(\ref{iija}) we get
\begin{eqnarray}
	\nonumber
	\frac{\partial^2 \mathcal{F}}{\partial S_i\partial S_j}
	& = &
	2 (1-\delta_{js}) \int_{\gamma}  \Log(z-z_i)^2 \rmd \phi_j(z_+)\\
	\label{iij2}
	&   & {}+ \frac{1}{\pi\rmi} \int_{\gamma}  \Log(z-z_i)^2\rmd \Omega_0(z_+),
	           \quad z_i\in\gamma_i.
\end{eqnarray}
Consider now the first integral in the right-hand side of~(\ref{iij2}) and denote
\begin{equation}
	F_j(u)  = \int_{\gamma}  \Log(z-u)^2 \rmd \phi_j(z_+), \quad u\in \gamma.
\end{equation}
It follows from~(\ref{nho}) that
\begin{equation}
	\rmd \phi_j(z_+) = -\rmd \phi_j(z_-),\quad z\in\gamma,
\end{equation}
which implies that
\begin{eqnarray}
	\nonumber
	F_j(u) & = & \int_{\gamma}  (\log(z_+-u)+\log(z_--u))\rmd \phi_j(z_+)\\
	\label{deff}
                  & = &-\int_{A}  \log(z-u)\rmd \phi_j(z),
\end{eqnarray}
where $A=A_1+\cdots+A_s$ is the sum of the contours in figure~\ref{fig:ai}. Therefore
\begin{equation}
	F'_j(u) = \int_{A}  \frac{\rmd \phi_j(z)}{z-u} = 0,\quad u\in\gamma,
\end{equation}
because the integrand is analytic outside $A$ and has residue zero at $\infty$.
Therefore the functions $F_j(u)$ are constant on any connected piece of $\gamma$.
Moreover, equation~(\ref{deff}) shows that the functions $F_j(u)$ are analytic in
$\mathbb{C}\setminus \Gamma$ and
\begin{equation}
	\label{fpr}
	F'_j(u) =  \int_{A}  \frac{\rmd \phi_j(z)}{z-u}
	           = -2\pi \,\rmi \, \frac{p_j(u)}{w(u)},\quad u\notin\gamma.
\end{equation}
Since
\begin{equation}
	\int_{A}  \rmd \phi_j(z)=0,
\end{equation}
it also follows from~(\ref{deff}) that $F_j(u)=\mathcal{O}(1/u)$ as $u\rightarrow\infty$.
Therefore, using the standard definition of the Abelian integrals
\begin{equation}
	\phi_j(u) = \int_{\infty_1}^u \rmd \phi_j(z)
\end{equation}
and equation~(\ref{fpr}), we may write
\begin{equation}
	F_j(u) = -2\pi\rmi \phi_j(u),\quad u\in\mathbb{C}\setminus \Gamma,
\end{equation}
and by continuity we conclude that
\begin{equation}
	\label{idfj}
	F_j(z_i) = -2\pi\rmi\phi_j(a_i^+),\quad z_i\in\gamma_i.
\end{equation}

The analysis of the second term in equation~(\ref{iij2}) is similar. We denote
\begin{equation}
	F(u) = \int_{\gamma}  \Log(z-u)^2\rmd \Omega_0(z_+),\quad u\in \gamma.
\end{equation}	
From (\ref{deo}) we deduce that
\begin{equation}
	\rmd\Omega_0(z_+) = -\rmd \Omega_0(z_-),\quad z\in\gamma,
\end{equation}
and the same arguments used for $F_j(u)$ show that $F(u)$ is constant on any connected piece of $\gamma$,
and that the derivative $F'(u)$ outside $\gamma$ is
\begin{equation}
	\label{fpu}
	F'(u) = -2\pi\,\rmi \,\frac{P_0(u)}{w(u)},\quad u\notin\gamma.
\end{equation}
Taking into account that
\begin{equation}
	\int_{\gamma}  \rmd \Omega_0(z) = -\pi\rmi,
\end{equation}
if we define the the Abelian integral $\Omega_0(u)$ as
\begin{equation}
	\Omega_0(u)
	=
	\lim_{u_0\rightarrow\infty} \left(\int_{u_0}^u \rmd \Omega_0(z)+\log u_0\right),
\end{equation}
and use equation~(\ref{fpu}), we find
\begin{equation}
	F(u) = -2\pi\rmi\Omega_0(u),\quad u\in\mathbb{C}\setminus \Gamma.
\end{equation}
Therefore, by continuity, we finally get the result
\begin{equation}
	\label{idf}
	F(z_i) = -2\pi\rmi\Omega_0(a_i^+),\quad \forall z_i\in\gamma_i.
\end{equation}
Equation~(\ref{ab20}) is now a direct consequence of~(\ref{iij2}), (\ref{idfj}) and~(\ref{idf}).
%%%%%%%%%%%%%%%%%%%%%%%%%%%%%%%%%%%%%%%%%%%%%%%%%%%%%%%%%%%%
%%  APPENDIX  B %%%%%%%%%%%%%%%%%%%%%%%%%%%%%%%%%%%%%%%%%%%%%%%%%%
%%%%%%%%%%%%%%%%%%%%%%%%%%%%%%%%%%%%%%%%%%%%%%%%%%%%%%%%%%%%
\section*{B Coalescence of cut endpoints}
The identities (\ref{ido})--(\ref{kbar}) which describe the behavior of the differentials  $\rmd \Omega_0$
and $\rmd\phi_1,\ldots,\rmd\phi_{s-1}$ under coalescence of the endpoints of two cuts can be proved
using a method due to Tian~\cite{TI94}. For conciseness  we provide the proof for  $\rmd \Omega_0$ only.
This differential can be written as
 \begin{equation}
	\rmd \Omega_0(z) = \frac{P_0(z)}{w(z)}\,\rmd z,
\end{equation}
where $P_0(z)$ is a polynomial of degree $s-1$  which depends on the endpoints
$\mathbf{a}^{(s)} = (a_1^{\pm},\ldots,a_s^{\pm})$ and is uniquely characterized
by~(\ref{norddb}) and~(\ref{nordd}). Let us first prove that~(\ref{dos}) implies
 \begin{equation}
 	\label{peo}
 	P_0^{(s)}(z;\mathbf{a}^{(s)})\Big|_{z=\alpha} = 0.
\end{equation}
From (\ref{norddb}) we have
\begin{equation}
\oint_{A_{m-1}}  \frac{P_0^{(s)}(z;\mathbf{a}^{(s)})}{w^{(s)}(z;\mathbf{a}^{(s)})}\rmd z=0,
\end{equation}
so that 
\begin{eqnarray*}
 \nonumber
 0 & = & \int_{a_{m-1}^{-}}^{a_{m-1}^{+}}
            \left.\frac{P_0^{(s)}(z;\mathbf{a}^{(s)})}{w^{(s)}(z;\mathbf{a}^{(s)})}\right|_{a_{m-1}^{(s)+}=a_m^{(s)-}=\alpha}
             \rmd z\\
   & = & \int_{a_{m-1}^{-}}^{\alpha}
             \frac{\left.P_0^{(s)}(z;\mathbf{a}^{(s)})\right|_{a_{m-1}^{(s)+}=a_m^{(s)-}=\alpha}
             }{
             (z-\alpha)\,w^{(s-1)}(z;\mathbf{a}^{(s-1)})}\rmd z,
\end{eqnarray*}
which implies (\ref{peo}).  Thus, for a coalescence of the type~(\ref{dos}) the function
\begin{equation}
	\frac{P_0^{(s)}(z;\mathbf{a}^{(s)})}{z-\alpha}
\end{equation}
determines a polynomial $P_0^{(s-1)}(z;\mathbf{a}^{(s-1)})$ of degree $s-2$ and
$\rmd \Omega_0^{(s)}$ reduces to
\begin{equation}
	\rmd \Omega_0^{(s)}(z) = \frac{P_0^{(s-1)}(z)}{w^{(s-1)}(z)}\,\rmd z.
\end{equation}
This expression determines the differential $\rmd \Omega_0^{(s-1)}(z)$ and therefore~(\ref{ido}) follows.
%%%%%%%%%%%%%%%%%%%%%%%%%%%%%%%%%%%%%%%%%%%%%%%%%%%%%%%%%%%%
%%  BIB %%%%%%%%%%%%%%%%%%%%%%%%%%%%%%%%%%%%%%%%%%%%%%%%%%%%%%%
%%%%%%%%%%%%%%%%%%%%%%%%%%%%%%%%%%%%%%%%%%%%%%%%%%%%%%%%%%%%
\providecommand{\href}[2]{#2}\begingroup\raggedright\endgroup
%%%%%%%%%%%%%%%%%%%%%%%%%%%%%%%%%%%%%%%%%%%%%%%%%%%%%%%%%%%%
%%  THE END %%%%%%%%%%%%%%%%%%%%%%%%%%%%%%%%%%%%%%%%%%%%%%%%%%%%
%%%%%%%%%%%%%%%%%%%%%%%%%%%%%%%%%%%%%%%%%%%%%%%%%%%%%%%%%%%%

\begin{thebibliography}{10}

\bibitem{CA01}
F.~Cachazo, K.~Intriligator, and C.~Vafa, {\it A large {$N$} duality via a
  geometric transition},  {\em Nuc.\ Phys.\ B} {\bf 603} (2001) 3.

\bibitem{DI02}
R.~Dijkgraaf and C.~Vafa, {\it Matrix models, topological strings, and
  supersymmetric gauge theories},  {\em Nuc.\ Phys.\ B} {\bf 644} (2002) 3.

\bibitem{DI022}
R.~Dijkgraaf and C.~Vafa, {\it On geometry and matrix models},  {\em Nuc.\
  Phys.\ B} {\bf 644} (2002) 21.

\bibitem{HE07}
J.~J. Heckman, J.~Seo, and C.~Vafa, {\it Phase structure of a brane/anti-brane
  system at large {$N$}},  {\em J. High Energy Phys.} {\bf 07} (2007) 073.

\bibitem{BI05}
A.~Bilal and S.~Metzger, {\it Special geometry of local {C}alabi-{Y}au
  manifolds and superpotentials from holomorphic matrix models},  {\em J. High
  Energy Phys.} {\bf 08} (2005) 097.

\bibitem{FE03}
F.~Ferrari, {\it Quantum parameter space in super {Y}ang-{M}ills. {II}},  {\em
  Phy.\ Lett.\ B} {\bf 557} (2003) 290.

\bibitem{CA03}
F.~Cachazo, N.~Seiberg, and E.~Witten, {\it Phases of {$N=1$} supersymmetric
  gauge theories and matrices},  {\em J.\ High Energy Phys.} {\bf 03} (2003)
  042.

\bibitem{HE08}
J.~J. Heckman and C.~Vafa, {\it Geometrically induced phase transitions at
  large {$N=1$}},  {\em J.\ High Energy Phys.} {\bf 04} (2008) 052.

\bibitem{MA10}
M.~Mari{\~n}o, S.~Pasquetti, and P.~Putrov, {\it Large {$N$} duality beyond the
  genus expansion},  {\em J.\ High Energy Phys.} {\bf 10} (2010) 074.

\bibitem{AL10}
G.~\'Alvarez, L.~{Mart\'{\i}nez Alonso}, and E.~Medina, {\it Phase transitions
  in multi-cut matrix models and matched solutions of {W}hitham hierarchies},
  {\em J. Stat.\ Mech.\ Theory Exp.} (2010) 03023.

\bibitem{BE95}
K.~Becker, E.~Becker, and A.~Strominger, {\it Fivebranes, membranes and
  non-perturbative string theory},  {\em Nuc.\ Phys.\ B} {\bf 456} (1995) 130.

\bibitem{KL96}
A.~Klemm, W.~Lerche, P.~Mayr, C.Vafa, and N.~Warner, {\it Self-dual strings and
  {$N=2$} supersymmetric field theory},  {\em Nuc.\ Phys.\ B} {\bf 477} (1996)
  746.

\bibitem{SH99}
A.~D. Shapere and C.~Vafa, ``{BPS} structure of {A}rgyres-{D}ouglas
  superconformal theories.'' arXiv:hep-th/9910182v2.

\bibitem{GU00}
S.~Gukov, C.~Vafa, and E.~Witten, {\it {CFT's} from {C}alabi-{Y}au four-folds},
   {\em Nuc.\ Phys.\ B} {\bf 584} (2000) 69.

\bibitem{GU00err}
S.~Gukov, C.~Vafa, and E.~Witten, {\it Erratum},  {\em Nuc.\ Phys.\ B} {\bf
  608} (2001) 477.

\bibitem{FE04}
G.~Felder and R.~Riser, {\it Holomorphic matrix integrals},  {\em Nuc.\ Phys.\
  B} {\bf 691} (2004) 251.

\bibitem{FE03b}
F.~Ferrari, {\it On exact superpotentials in confining vacua},  {\em Nuc.\
  Phys.\ B} {\bf 648} (2003) 161.

\bibitem{FE03d}
F.~Ferrari, {\it Quantum parameter space and double scaling limits in {$N=1$}
  super {Y}ang-{M}ills theory},  {\em Phys.\ Rev.\ D} {\bf 67} (2003) 085013.

\bibitem{SI95}
Y.~Sibuya, {\em Global Theory of a Second Order Linear Ordinary Differential
  Equation with a Polynomial Coefficient}.
\newblock North-Holland, 1975.

\bibitem{BE11}
M.~Bertola, {\it Boutroux curves with external field: equilibrium measures
  without a variational problem},  {\em Analysis and Math.\ Phys.} {\bf 1}
  (2011) 167.

\bibitem{GR80}
D.~Gross and E.~Witten, {\it Possible third-order phase transition in the
  large-{N} lattice gauge theory},  {\em Phys.\ Rev.\ D} {\bf 21} (1980) 446.

\bibitem{BL03}
P.~Bleher and B.~Eynard, {\it Double scaling limit in random matrix models and
  a nonlinear hierarchy of differential equations. {R}andom matrix theory.},
  {\em J.\ Phys.\ A: Math.\ Gen.} {\bf 36} (2003) 3085.

\bibitem{BR78}
E.~Br{\'e}zin, C.~Itzykson, G.~Parisi, and J.~B. Zuber, {\it Planar diagrams},
  {\em Commun.\ Math.\ Phys.} {\bf 59} (1978) 35.

\bibitem{EY06}
B.~Eynard, {\it Universal distribution of random matrix eigenvalues near the
  birth of a cut},  {\em J. Stat.\ Mech.\ Theory Exp.} (2006) P07005.

\bibitem{SE94}
N.~Seiberg and E.~Witten, {\it Monopole condensation, and confinement in
  {$N=2$} supersymmetric {Y}ang-{M}ills theory},  {\em Nuc.\ Phys.\ B} {\bf
  426} (1994) 19.

\bibitem{CA07}
N.~Caporaso, L.~Griguolo, M.~Mari{\~n}o, S.~Pasquetti, and D.~Seminara, {\it
  Phase transitions, double-scaling limit and topological strings},  {\em
  Phys.\ Rev.\ D} {\bf 75} (2007) 046004.

\bibitem{KL10}
A.~Klemm, M.~Mari{\~{n}}o, and M.~Rauch, {\it Direct integration and
  non-perturbative effects in matrix models},  {\em J. High Energy Phys.} {\bf
  10} (2010) 004.

\bibitem{TI94}
F.~R. Tian, {\it The {W}hitham type equations and linear overdetermined systems
  of {E}uler-{P}oisson-{D}arboux type},  {\em Duke.\ Math.\ J.} {\bf 74} (1994)
  203.

\end{thebibliography}
\end{document}